\documentclass[11pt,a4paper]{article}
\pdfoutput=1
\usepackage{jheppub} 
\usepackage{graphics,graphicx}
\usepackage{amsfonts, amsmath, amsthm, amssymb,latexsym, bm}
\usepackage[retainorgcmds]{IEEEtrantools}
\usepackage[title,titletoc]{appendix}
\makeatletter

\newcommand{\Rmnum}[1]{\expandafter\@slowromancap\romannumeral #1@}
\newcommand{\bra}[1]{\langle #1|}
\newcommand{\ket}[1]{|#1\rangle}

\newcommand{\Hilbert}{\mathcal{H}}

\newcommand{\pvec}{\mathbf{p}}
\newcommand{\qvec}{\mathbf{q}}
\newcommand{\kvec}{\mathbf{k}}
\newcommand{\ud}[2][{}]{\!\mathrm{d}^{#1}#2\;}
\newcommand{\udpi}[2][{}]{\!\frac{\mathrm{d}^{#1}#2}{(2\pi)^{#1}}\;}
\newcommand{\Itensorup}[4]{\eta^{#1#3}\eta^{#2#4} + \eta^{#1#4}\eta^{#2#3} - \eta^{#1#2}\eta^{#3#4}}
\newcommand{\Itensordown}[4]{\eta_{#1#3}\eta_{#2#4} + \eta_{#1#4}\eta_{#2#3} - \eta_{#1#2}\eta_{#3#4}}
\DeclareMathOperator{\imaginary}{Im}
\DeclareMathOperator{\real}{Re}

\renewcommand{\vec}[1]{\mathbf{#1}}

\makeatother

\title{Construction of an asymptotic S matrix for perturbative quantum gravity}

\author{John Ware, Ryo Saotome and Ratindranath Akhoury}

\affiliation{Michigan Center for Theoretical Physics, Randall Laboratory of Physics, University of Michigan, Ann Arbor, MI 48109-1120, USA}

\emailAdd{johnware@umich.edu, rsaotome@umich.edu, akhoury@umich.edu}

\abstract{The infrared behavior of perturbative quantum gravity is studied using the method developed for QED by Faddeev and Kulish.  The operator describing the asymptotic dynamics is derived and used to construct an IR-finite S matrix and space of asymptotic states.  All-orders cancellation of IR divergences is shown explicitly at the level of matrix elements for the example case of gravitational potential scattering. As a practical application of the formalism, the soft part of a scalar scattering amplitude is related to the gravitational Wilson line and computed to all orders.}

\arxivnumber{}

\begin{document}
\maketitle

\flushbottom

\section{Introduction}

The problem of infrared divergences in perturbative quantum gravity was first investigated by Weinberg \cite{Weinberg}, who showed the exponentiation of soft divergences for a class of ladder and crossed ladder diagrams in the eikonal approximation. Subsequently, the cancellation of soft divergences was also demonstrated at one loop in inclusive cross sections for graviton-graviton scattering \cite{Donoghue}. Work so far, however, has relied on the Bloch-Nordsieck method \cite{BN}, in which soft divergences cancel only when the transition probability is summed over nearly degenerate final state processes. Because the cancellation occurs at the level of the cross section, this leaves us without a (finite) definition of the S matrix in four dimensions. This shortcoming is of relevance for investigations involving the gravitational S matrix, which has attracted particular interest both in the context of black hole production and decay \cite{Giddings1, 'tHooft} and in the perturbative evaluation of gravitational amplitudes (for reviews see \cite{amplitudes, amplitudes2, amplitudes3}). The lack of an S matrix obscures the internal logic of the theory and may impede future progress.

Luckily, we can draw on our experience with another theory -- QED -- in which a long-range interaction leads to infrared divergences. Faddeev and Kulish \cite{FK} constructed an IR-finite S matrix for QED as early as 1970. Their method considered the interaction potential in the limit of long times. Due to the long range of the electromagnetic force, a part of this potential survives in the limit $|t|\rightarrow\infty$. Asymptotic (scattering) states are therefore governed not by the free time-evolution operator $\exp(-iH_{0}t)$ but by a different operator $U_{as}$. 

The resulting matrix elements may be understood in two equivalent ways \cite{Contopanagos}. We can use the usual (Dyson) S matrix but define a new space of asymptotic states in which every charged particle comes `dressed' with a coherent cloud of soft photons. This picture allows for calculations using the familiar Feynman rules, but it obscures the particle content of the theory, since the asymptotic space admits no irreducible massive representations of the Poincar\'e group. Alternatively, we can continue to use the Fock space as the space of scattering states if the S matrix is modified from its usual form. From this point of view, the Hilbert space and therefore the particle content are unaffected.

Can the same technique be taken over to gravitation? Until recently, the main obstactle to extending the Faddeev-Kulish program to perturbative quantum gravity has been the lack of a complete characterization of IR divergences in that theory. This problem has now \cite{Coll} been solved. 
Despite the highly nonlinear nature of gravity, the IR behavior is remarkably simple. There are no collinear divergences, and soft divergences arise only in ladder and crossed ladder diagrams constructed from three-point vertices, i.e. from linearized interactions. These results make it possible to carry out the Faddeev-Kulish procedure for gravity without treating an infinite tower of derivative interactions.

In this paper, we derive the operator $U_{as}$ of asymptotic dynamics for perturbative quantum gravity and use it to construct an IR-finite S matrix (equivalently, a space of asymptotic states). We show the cancellation of soft divergences to all orders when our S matrix is used in one-particle potential scattering.  

In section \ref{sec:power} we briefly review the power-counting argument of \cite{Coll}, which establishes that we need only consider linearized interactions.
In section \ref{sec:operator}, we derive the operator that determines the gravitational asymptotic dynamics and show that it serves to cancel the divergent `Coulomb' phase factors that arise in S matrix elements. 
In the following section, \ref{sec:asymptotic}, we define the asymptotic S matrix and construct the asymptotic graviton and scalar field operators. We show explicitly that the asymptotic operators create matter particles along with the associated classical gravitational fields. 
Next, in section \ref{sec:coherent}, we discuss the gauge and Lorentz transformation properties of the gravitational coherent states.  We then explicitly construct the physical asymptotic states relevant to one-particle gravitational potential scattering. 

In section \ref{sec:cancellation}, following the work of Chung in QED \cite{Chung} and treating potential scattering explicitly as a model case, we show that our S matrix yields IR-finite matrix elements to all orders. This establishes one of the main advantages of the Faddeev-Kulish method over the approach used in previous work including that of Weinberg \cite{Weinberg} and Donoghue \cite{Donoghue}, in which no finite S matrix elements can be defined and infrared cancellations are seen only at the level of inclusive cross sections.
In section \ref{sec:scattering}, we use the results of the previous section to consider the infrared graviton corrections to all orders to an exclusive hard scattering process. The virtual corrections exponentiate and the collinear divergences cancel when corrections to all legs are included. Our results here are in agreement with those of \cite{Schnitzer1} (see also \cite{Schnitzer2}) and show clearly the relevance of gravitational Wilson lines to the soft part of graviton scattering in perturbative quantum gravity. 
The final section summarizes our conclusions. Certain technical details are relegated to an appendix.

\section{Gravitational collinear and soft divergences}
\label{sec:power}

In this section, we briefly review the results of \cite{Coll}, in which we characterize, to all orders, the diagrams contributing to infrared divergences in perturbative gravity. In subsection \ref{sec:jet} we consider collinear divergences; in subsection \ref{sec:soft} we turn to soft divergences. We find that only the latter are directly relevant to the Faddeev-Kulish construction.  We also show that it suffices, for discussions of infrared behavior, to treat the gravitational field only to linear order in interactions.

\subsection{Jet power counting}
\label{sec:jet}

\begin{figure}
\begin{center}
\includegraphics[width=2.5in]{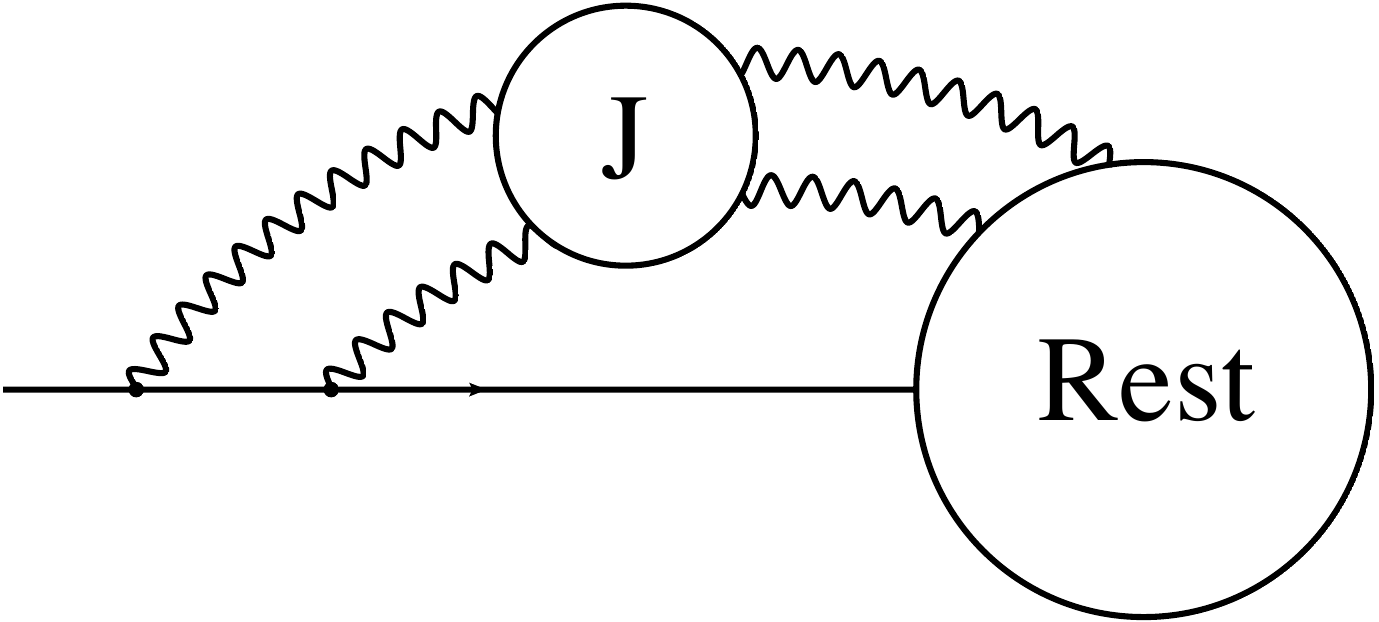}
\end{center}
\caption{An diagram with a graviton jet attached to an external leg.}
\label{BaseFig}
\end{figure}

Consider the addition of a graviton jet attached to a massless on-shell line with momentum $p$ in an arbitrary diagram (see figure \ref{BaseFig}). The addition of such a graviton jet will require us to integrate over all independent loop momenta in the jet. In each jet loop integral, we can make the change of variables
\begin{align}
\int d^{4}l\sim\int dl_{\bot}^{2}dl_{+}dl_{-},
\end{align}
where $l_{\bot}^{2}$ are the two components of the loop momentum $l$ transverse to $p$ and $l_{\pm}$ is defined as $\frac{1}{\sqrt{2}}(l_{0}\pm \vec{l}\cdot\hat{\vec{p}})$. We define the collinear degree of divergence $\gamma_{co}$ as the total number of factors of vanishingly small variables, called normal variables, that scale in the same way as $l_{-}$. 

We can boost to a frame in which $l_{-}$ and $l_{\bot}^{2}$ are small. Therefore, when $l$ and $p$ are collinear, the jet loop contributes two normal variables in the numerator. Note that $l_{-}$, $l_{\bot}^{2}$, and $l^{2}$ all scale together, since $l^{2}=l_{+}l_{-}-l_{\bot}^{2}$ and $l_{+}$ does not vanish. Since each propagator in the jet contributes a factor of $l^{2}$ in the denominator, each jet line will will subtract one from the collinear degree of divergence. The only other additional factors of normal variables come from vertex numerator factors. So we can write the collinear degree of divergence $\gamma_{co}$ as
\begin{align}
\gamma_{co}=2L_{J}-N_{J}+N_{num},
\label{base}
\end{align}
where $L_{J}$ is the number of loops in the jet, $N_{J}$ the number of jet lines, and $N_{num}$ the total number of normal variables in the numerator arising from vertices.

\begin{figure}
\begin{center}
\includegraphics[width=2.5in]{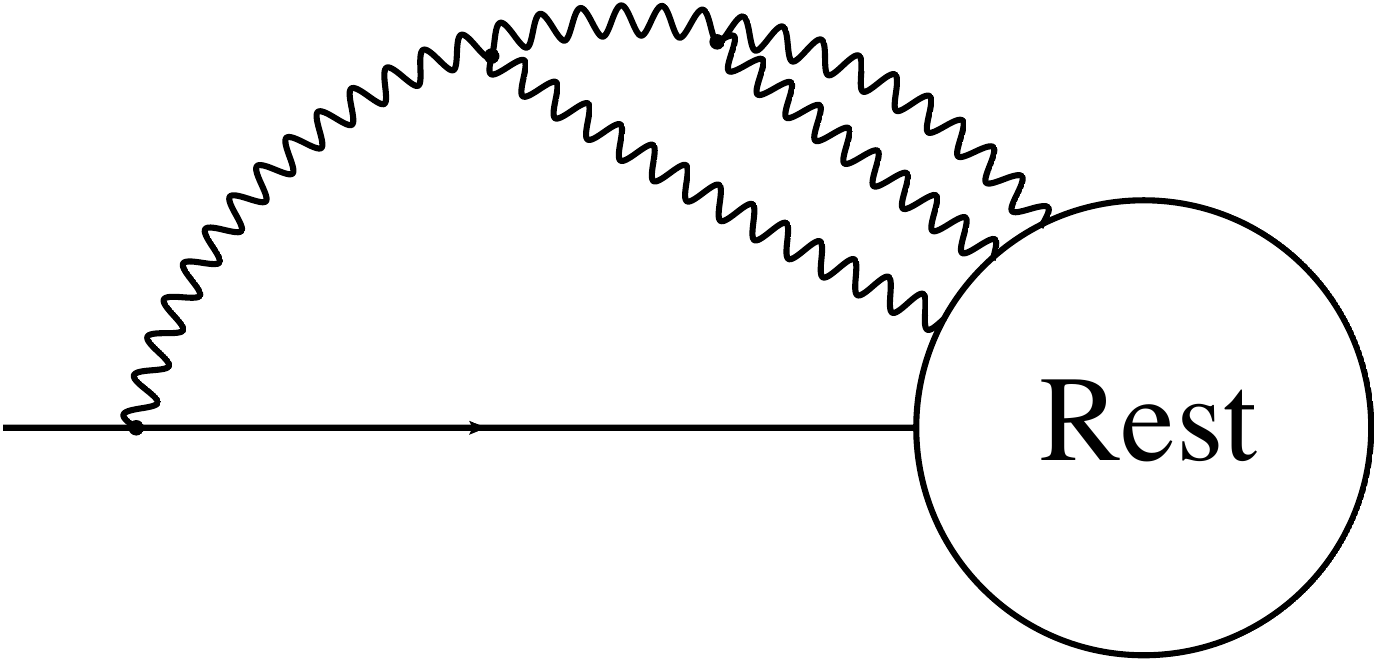}
\end{center}
\caption{A diagram with no internal jet loops and only three point vertices.}
\label{Tree}
\end{figure}

In \cite{Coll}, we show that gravitational collinear divergences can arise only when a jet has only three point vertices and no internal jet loops. An example is shown in figure \ref{Tree}. Note that jets of this kind always have $2L_{J}=N_{J}$.

In the case that such a jet contains only gravitational vertices, $N_{num}=0$. This seems surprising because each vertex within the jet produces two factors of collinear momenta. However, for every vertex where the jet attaches to a non-collinear line (labeled ``rest" in figure \ref{BaseFig}), two factors of collinear momenta are contracted with two factors of non-collinear momenta. But for diagrams with no internal jet loops and only three point vertices, there are the same number of collinear jet vertices as there are vertices that connect the jet to ``rest."  So $N_{num}=0$. Therefore jets of this kind have $\gamma_{co}=0$, indicating a logarithmic collinear divergence.

We have seen that individual diagrams can contain collinear divergences. What happens to this collinear divergence when we include all the topologically distinct diagrams contributing to a particular process? In \cite{Coll} it was shown (using the gravitational Ward identity) that when all possible additions of jets to an arbitrary diagram are summed over, the collinear divergences cancel. This cancellation, which is thus seen to be a consequence of momentum conservation and the on-shell condition, was first observed in the context of the eikonal approximation in \cite{Weinberg}. The analysis of \cite{Coll} relies on no approximations and was confirmed in \cite{Beneke}.

Since the collinear divergences already cancel at the level of matrix elements, they will play no role in the Fadeev-Kulish construction.

\subsection{Soft power counting}
\label{sec:soft}

\begin{figure}
\begin{center}
\includegraphics[width=2.5in]{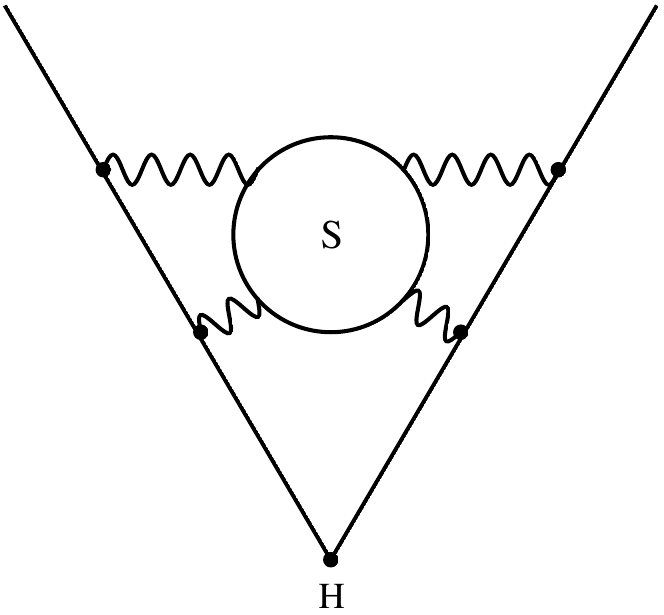}
\end{center}
\caption{Arbitrary diagram with virtual soft graviton corrections to a hard vertex. The finite momentum lines are drawn with a solid line.}
\label{BaseSoft}
\end{figure}

We now review the power-counting procedure for soft divergences. Consider soft graviton corrections to a hard vertex, as in figure \ref{BaseSoft}. Note that the spin of the hard lines is immaterial, since soft gravitons couple independently of spin \cite{Weinberg2}.

\begin{figure}
\begin{center}
\includegraphics[width=2.5in]{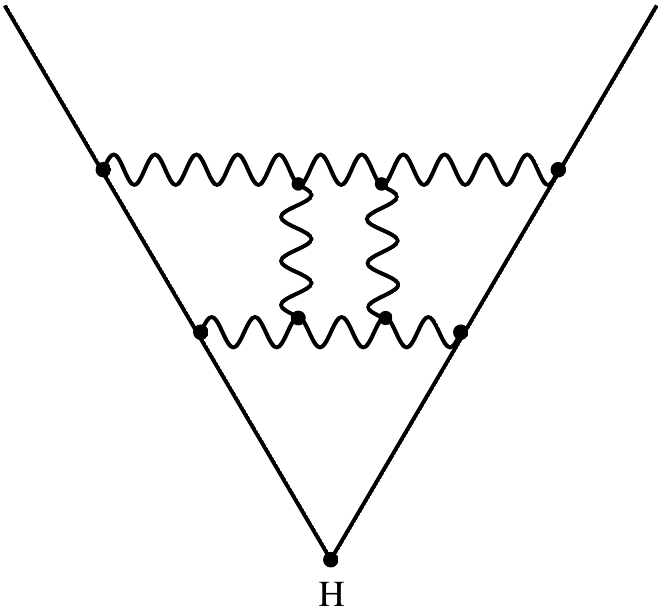}
\end{center}
\caption{An example of the type of diagram discussed in section \ref{sec:soft}. Here $L_{S}=4$, $N_{S}=8$, and $N_{E}$=4.}
\label{SoftEx}
\end{figure}

For soft divergences, all four components of the graviton loop 4-momentum are normal variables. Therefore, each graviton loop now adds four and each virtual graviton line subtracts two from the soft degree of divergence. Each virtual finite momentum line will subtract one from the soft degree of divergence, since the denominator of the corresponding propagator is linear in graviton loop momenta in the eikonal regime. So we have
\begin{align}
\gamma_{soft}=4L_{S}-2N_{S}-N_{E}+N_{sn},
\label{soft1}
\end{align}
where $L_{S}$ is the number of soft loops in S, $N_{S}$ is the number of soft graviton lines in S, $N_{E}$ is the number of virtual finite momentum lines in the diagram, and $N_{sn}$ is the contribution of soft normal variables to the numerator from the vertices. For clarification of these quantities, see the example given in figure \ref{SoftEx}.

\begin{figure}
\begin{center}
\includegraphics[width=2.5in]{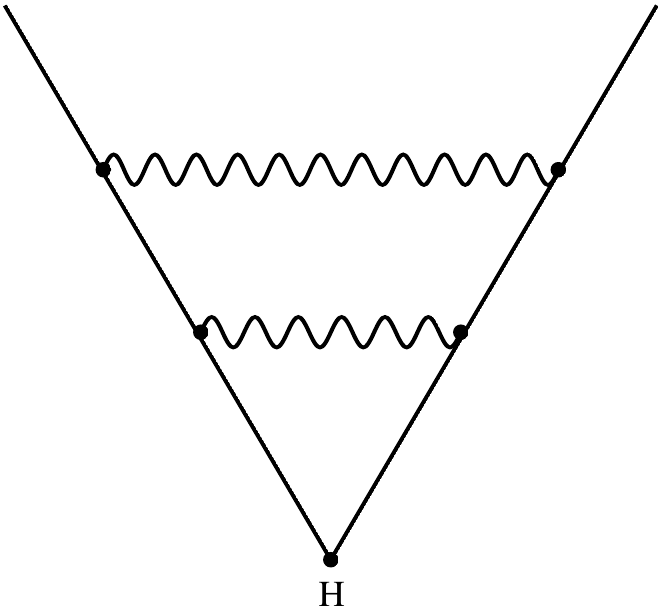}
\end{center}
\caption{An example of a diagram with only three point couplings to hard lines that will lead to a soft divergence.}
\label{Soft}
\end{figure}

In \cite{Coll} it was shown that soft divergences can occur in 1PI diagrams only if all soft lines attach only to finite momentum lines. An example of such a diagram is shown in figure \ref{Soft}. In such diagrams, $\gamma_{soft}$ always vanishes, since $L_{S}=N_{S}$, $N_{E}=2N_{S}$, and $N_{sn}=0$. Diagrams of this kind may therefore contain logarithmic soft divergences. In general, such divergences cancel only at the level of the cross section, and only if we allow the emission or absorption of an arbitrary number of real soft gravitons. 

We have shown in this section that vertices involving more than one soft graviton line do not contribute to soft divergences. As a result, interactions at quadratic order or higher in the gravitational field will play no role in the IR behavior of the theory.  In particular, we can safely carry out the Faddeev-Kulish procedure in perturbative quantum gravity using only linearized interactions (three-point vertices). 

It is worth pointing out that the absence of collinear divergences and the simple structure of soft divergences in perturbative quantum gravity makes this theory more tractable in the infrared regime than even QED. For this reason, the Faddeev-Kulish approach is on a firmer foundation here than in the case of massless gauge theories, which are fraught with collinear divergences. 

\section{Finding the asymptotic operator}
\label{sec:operator}

As discussed in the previous section, in order to find the correct asymptotic states for perturbative quantum gravity, we need only consider terms up to linear order in the gravitational field. Since soft gravitons do not see spin, the asymptotic behavior should be independent of the spin of the matter. For simplicity, we consider a single scalar field coupled to gravity. The relevant Lagrangian is
\begin{align}
\mathcal{L}=\sqrt{g}(-R-\frac{1}{2}\partial_{\mu}\phi g^{\mu \nu}\partial_{\nu}\phi-\frac{1}{2}m^{2}\phi^{2})+\mathcal{L}_{g.f.},
\label{Lag}
\end{align}
where we have taken $\kappa^{2}=16\pi G=c=\hbar=1$ and omitted the ghost Lagrangian, which does not contribute to infrared divergences \cite{deWitt}. Expanding about the flat metric $g_{\mu \nu} = \eta_{\mu\nu}+h_{\mu\nu}$ to linear order in $h$,
\begin{IEEEeqnarray}{rCl}
g^{\mu\nu} & = & \eta^{\mu\nu}-h^{\mu\nu},
\nonumber
\\
\sqrt{g} & = & 1+\frac{1}{2}h^{\mu}_{\mu},
\\
\nonumber
R & = & \partial^{2}h^{\mu}_{\mu}-\partial_{\mu}\partial_{\nu}h^{\mu\nu}.
\end{IEEEeqnarray}
We choose the gauge fixing term $\mathcal{L}_{g.f.}=\frac{1}{2}C^{2}$, $C^{\mu}=\partial_{\nu}h^{\mu\nu}-\frac{1}{2}\partial^{\mu}h^{\nu}_{\nu}$.
Substituting these definitions into \eqref{Lag} and dropping total derivatives, we have, to linear order in $h$ in the interaction potential,
\begin{IEEEeqnarray}{rCl}
\mathcal{L} &=& -\frac{1}{2}\eta^{\mu\nu}\partial_{\mu}\phi\partial_{\nu}\phi-\frac{1}{2}m^{2}\phi^{2}-\frac{1}{4}\partial_{\alpha}h_{\mu\nu}\partial^{\alpha}h^{\mu\nu}+\frac{1}{8}\partial_{\alpha}h^{\mu}_{\mu}\partial^{\alpha}h^{\nu}_{\nu}-\mathcal{V}_{grav},
\\
\nonumber
\mathcal{V}_{grav} &=& -\frac{1}{2}h^{\mu\nu}\partial_{\mu}\phi\partial_{\nu}\phi+\frac{1}{4}h^{\alpha}_{\alpha}\eta^{\mu\nu}\partial_{\mu}\phi\partial_{\nu}\phi+\frac{1}{4}h^{\alpha}_{\alpha}m^{2}\phi^{2}.
\end{IEEEeqnarray}
The graviton propagator corresponding to this Lagrangian is
\begin{equation}
\frac{i}{2}[\eta^{\alpha\gamma}\eta^{\beta\delta}+\eta^{\beta\gamma}\eta^{\alpha\delta}-\eta^{\alpha\beta}\eta^{\gamma\delta}]\frac{1}{l^{2}+i\epsilon}.
\end{equation}

We are interested in the asymptotic behavior of the interaction Lagrangian $V = \int\ud[3]{x}\mathcal{V}_{grav}$, which will determine the asymptotic Hamiltonian. In order to see what terms remain when we take the asymptotic limit $|t|\rightarrow\infty$, first expand the field operators in terms of creation and annihilation operators:

\begin{IEEEeqnarray}{rCl}
\phi({\bf{x}},t) &=& \int \!\frac{\mathrm{d}^{3}p}{(2\pi)^{3/2}}\frac{1}{\sqrt{2p^{0}}}\left(b(p)e^{ip\cdot x}+b^{\dag}(p)e^{-ip\cdot x}\right),
\label{comm1}\\
h_{\mu\nu}({\bf{x}},t) &=& \int \!\frac{\mathrm{d}^{3}k}{(2\pi)^{3/2}}\frac{1}{\sqrt{2k^{0}}}\left(a_{\mu\nu}(k)e^{ik\cdot x}+a_{\mu\nu}^{\dag}(k)e^{-ik\cdot x}\right)
\label{comm2}
\end{IEEEeqnarray}
with commutation relations
\begin{IEEEeqnarray}{rCl}
[b(p),b^{\dag}(p')] &=& \delta^{(3)}({\bf p-p'}),
\\
{[a_{\mu\nu}(k),a_{\sigma\lambda}^{\dag}(k')]} &=& (\eta_{\mu\sigma}\eta_{\nu\lambda}+\eta_{\mu\lambda}\eta_{\nu\sigma}-\eta_{\mu\nu}\eta_{\sigma\lambda})\delta^{(3)}({\bf k-k'}).
\end{IEEEeqnarray}
Equations \eqref{comm1} and \eqref{comm2} are both expressed in the interaction representation, which coincides with the Heisenberg representation for the free field case. When we substitute these expressions into $V$, the spatial integral will produce a delta function $\delta^{(3)}(\pvec\pm\qvec-\kvec)$ relating the momenta $\vec{p}$ and $\vec{q}$ of the scalars and $\vec{k}$ of the graviton. The resulting expression will contain terms of two types: 1) those with two creation or two annihilation operators of the scalars and 2) those with one creation and one annihilation operator of the scalars. Terms of the first type will carry the exponential time dependence
\begin{align}
&\exp(-i(p^{0}+q^{0}\pm k^{0})t)
\nonumber \\
=&\exp(-i(\sqrt{{\bf{p}}^{2}-m^{2}}+\sqrt{({\bf{p+k}})^{2}-m^{2}}\pm k^{0})t).
\label{firsttype}
\end{align}
For large $t$, this factor becomes highly oscillatory, so terms of type \eqref{firsttype} average to zero in the integration over the remaining momenta $\vec{p}$ and $\vec{k}$.  Terms of the second type carry the time dependence
\begin{align}
&\exp(-i(p^{0}-q^{0}\pm k^{0})t)
\nonumber \\
=&\exp(-i(\sqrt{{\bf{p}}^{2}-m^{2}}-\sqrt{({\bf{p+k}})^{2}-m^{2}}\pm k^{0})t).
\label{secondtype}
\end{align}
Due to the difference in sign, the argument of the exponential vanishes along the whole line $\vec{k}=0$, $\vec{p}$ arbitrary, supressing the oscillatory behavior. We therefore expect the integral to be dominated for large $t$ by the region of small $\vec{k}$. Neglecting terms of the type \eqref{firsttype} compared to those of type \eqref{secondtype} and treating $\vec{k}$ as small, the asymptotic potential becomes, for large $t$,
\begin{align}
V^{I}_{as}(t)=-\frac{1}{(2\pi)^{3/2}}\int\!\frac{\mathrm{d}^{3}p\,\mathrm{d}^{3}k}{2p^{0}\sqrt{2k^{0}}}p^{\mu}p^{\nu}(a_{\mu\nu}(k)e^{i\frac{k\cdot p}{p^{0}}t}+a^{\dag}_{\mu\nu}(k)e^{-i\frac{k\cdot p}{p^{0}}t})\rho(p),
\label{IP}
\end{align}
where $\rho(p)$ is the (unintegrated) number operator $b^{\dag}(p)b(p)$.

We seek to find the operator $U_{as}(t)$ that defines the asymptotic dynamics. This operator can be found by solving the Schrodinger equation,
\begin{align}
i\frac{d}{dt}U_{as}(t)=H_{as}(t)U_{as}(t),
\end{align}
where $H_{as}(t)=H_{0}+V_{as}(t)$, $H_{0}$ is the free field Hamiltonian, and $V_{as}(t)$ is \eqref{IP} in the Schrodinger representation. Paralleling the construction of the asymptotic operator in the QED case, we make the ansatz
\begin{align}
U_{as}(t)=e^{-iH_{0}t}Z(t),
\end{align}
which gives the following equation for $Z(t)$:
\begin{align}
i\frac{d}{dt}Z(t)=V^{I}_{as}(t)Z(t).
\label{Zeq}
\end{align}
The solution to an equation of the form \eqref{Zeq} is a time-ordered exponential, but because $V^{I}_{as}(t)$ commutes with its own commutator $Q(t_{1},t_{2})=[V^{I}_{as}(t_{1}),V^{I}_{as}(t_{2})]$, we can also write $Z(t)$ in the explicit form
\begin{align}
Z(t)=\exp\left(-i\int^{t}\ud{\tau}V^{I}_{as}(\tau)-\frac{1}{2}\int^{t}\ud{\tau}\int^{\tau}\ud{s}Q(\tau,s)\right).
\label{FormZ}
\end{align}
Note that when evaluating the integrals in \eqref{FormZ}, there should be no dependence on any constant finite time. This is because the wave packets $U_{as}(t)\Psi$ must behave classically for large $|t|$, and $V^{I}_{as}(t)$ does not commute with the momentum of the system for finite $t$ as it is linearly dependent on $a_{\mu\nu}(k)$ and $a^{\dag}_{\mu\nu}(k)$. So we must evaluate the integrals in \eqref{FormZ} using \cite{FK}:
\begin{align}
\int^{t}\ud{\tau}e^{is\tau}=\frac{1}{is}e^{ist}.
\end{align}
Using \eqref{R} and \eqref{Phi}, we can write the asymptotic operator as
\begin{align}
U_{as}(t)=e^{-iH_{0}t}e^{i\Phi (t)}e^{R(t)},
\end{align}
where
\begin{IEEEeqnarray}{rCl}
R(t) & = &-i\int^{t}\ud{\tau}V^{I}_{as}(\tau)
\label{R}
 \\
&=&\frac{1}{(2\pi)^{3/2}}\int\!\frac{\mathrm{d}^{3}p\,\mathrm{d}^{3}k}{2k\cdot p\sqrt{2k^{0}}}p^{\mu}p^{\nu}(a_{\mu\nu}(k)e^{i\frac{k\cdot p}{p^{0}}t}-a^{\dag}_{\mu\nu}(ke^{-i\frac{k\cdot p}{p^{0}}t})\rho(p),
\nonumber \\
\label{Phi}
\Phi(t)&= &\frac{i}{2}\int^{t}\ud{\tau}\int^{\tau}\ud{s}Q(\tau,s)
 \\
&=&-\frac{1}{32\pi}\int \ud[3]{p}\ud[3]{q}\!:\rho(p)\rho(q):\frac{2(p\cdot q)^{2}-m^{4}}{\sqrt{(p\cdot q)^{2}-m^{4}}}
\int^t\!\frac{\mathrm{d}\tau}{|\tau|}.
\nonumber
\end{IEEEeqnarray}

The role of the operator $e^{i\Phi (t)}$ is to cancel the divergent phase factor that would otherwise arise when acting $e^{-iH_{0}t}$ on states in $\mathcal{H}_{as}$. These divergent phase factors for both QED and linearized gravity were calculated by Weinberg \cite{Weinberg}. In an appendix we calculate this using dimensional regularization to regulate the IR divergences instead of introducing a graviton mass. We find that each pair of particles $m$, $n$ in the initial or final state produces the phase factor
\begin{align}
\phi_{mn}=\frac{i}{16\pi}\frac{m_nm_m(1+\beta_{mn}^{2})}{\beta_{mn}\sqrt{1-\beta_{mn}^{2}}}\left(\frac{1}{\epsilon}+\text{finite}\right),
\label{gravphi}
\end{align}
where $\beta_{mn}=\sqrt{1-\frac{m_n^2m_m^2}{(p_n\cdot p_m)^{2}}}$ is the relative velocity of the particles. 
In eq. \eqref{Phi} the infrared divergence in the phase operator is regulated by time. To show the cancellation of the phase eq. \eqref{gravphi}, we instead take $t\rightarrow\infty$ and regulate this divergence by dimensional regularization. The asymptotic phase operator eq. \eqref{Phi}, involves the integral
\begin{equation}
I(t)=\int^t \!\frac{\mathrm{d}\tau}{|\tau|},
\label{tauintegral}
\end{equation}
which diverges as $t\rightarrow\infty$.  We instead write
\begin{equation}
I(t\rightarrow\infty) = \int^\infty \!\frac{\mathrm{d}^{n-3}\tau}{|\tau|} = \frac{2\pi^{(n-3)/2}}{\Gamma((n-3)/2)}\int^\infty \ud{w}w^{n-5}.
\end{equation}
In this case, to regulate the divergence at the upper limit, we should take $n<4$, which gives
\begin{equation}
\frac{2\pi^{(n-3)/2}}{\Gamma((n-3)/2)}\frac{1}{n-4} = \frac{2}{4-n} + \text{IR finite}.
\end{equation}
Then the asymptotic phase operator is
\begin{equation}
\Phi(\infty) = -\frac{1}{16\pi}\int\ud[3]{p}\ud[3]{q} :\rho(p)\rho(q): \frac{2(p\cdot q)^2-m^4}{\sqrt{(p\cdot q)^2-m^4}} \frac{1}{\epsilon},
\end{equation}
where here $\epsilon = 4-n >0$.
This will produce, for each pair of particles in the state it acts on, the phase factor (in the case that all external particles have the same mass $m$)
\begin{equation}
-\frac{i}{16\pi}\frac{2(p_n\cdot p_m)^2-m^4}{\sqrt{(p_n\cdot p_m)^2-m^4}} \frac{1}{\epsilon} = -\frac{i}{16\pi}\frac{m^2(1+\beta_{mn}^2)}{\beta_{mn}\sqrt{1-\beta_{mn}^2}}\frac{1}{\epsilon}  = -\phi_{mn},
\end{equation}
which exactly cancels the divergent phase arising from internal soft graviton lines, eq. \eqref{gravphi}.

The remainder of our discussion will focus primarily on the operator $W(t) = e^{R(t)}$, which serves to associate a coherent cloud of soft gravitons to each hard matter particle.

\section{Asymptotic dynamics and asymptotic fields}
\label{sec:asymptotic}
The usual derivation of Dyson S matrix requires that the scattering states behave like states of the free theory as $t \rightarrow \pm \infty$. In the presence of long-range interactions mediated by massless fields, this condition fails, leading to infrared divergences. In particular, as we have shown, the Hamiltonian in the standard interaction picture does not switch off at large times. The Faddeev-Kulish procedure may be thought of as defining a new asymptotic interaction picture, the fields of which are related to the free fields by a (formally unitary) transformation. As discussed earlier, this transformation is generated by the asymptotic Hamiltonian described in the previous section and is given for large $t$ by
\begin{equation}
Z(t) = \mathrm{T}\exp\left[-i\int^{t} \ud{\tau} V_{as}(\tau)\right],
\end{equation}
where, for the case of perturbative quantum gravity,
\begin{equation}
V_{as}(t)=-\int\ud[3]{x} h^{\mu\nu}(t,\vec{x})T^{as}_{\mu\nu}(t,\vec{x}), \quad T^{as}_{\mu\nu} = \int\ud[3]{p} \frac{p_\mu p_\nu}{2 p_0} \rho(p) \delta^{(3)}(\vec{x}-t\pvec/p_0).
\end{equation}
It will be explicitly shown in the next section that the S matrix defined by
\begin{equation}
S_A = \lim_{t\rightarrow\infty} Z^{\dagger}(t)S_D Z(t),
\end{equation} 
where $S_D$ is the usual Dyson S matrix,
is free of infrared divergences to all orders. In this section we will analyze the nature of the asymptotic fields and their interpretations.
We first show by a unitary transformation of the free field $h_{\mu\nu}$ that $T_{\mu\nu}^{as}$ drives an asymptotic field $h_{\mu\nu}^{as}$:
\begin{equation}
h_{\mu\nu}^{as}(t,\vec{x}) = Z^\dagger(t) h_{\mu\nu}(t,\vec{x} Z(t).
\end{equation}
Recall that $Z(t)$ can be written
\begin{equation}
Z(t) = \exp\left[-i\int^t\ud{\tau} V_{as}(\tau)\right] \exp\left[-\frac{1}{2}\int^t\ud{\tau}\int^{\tau}\ud{s} [V_{as}(\tau),V_{as}(s)]\right].
\end{equation}
The commutator $[V_{as}(\tau),V_{as}(s)]$ itself commutes with $V_{as}$ and with $h_{\mu\nu}$, so it cancels in $Z^\dagger h_{\mu\nu} Z$.  This leaves
\begin{IEEEeqnarray}{rCl}
\nonumber
h^{as}_{\mu\nu}(t,\vec{x}) & = & \mathrm{\bar{T}}\exp\left[i\int^t\ud{\tau} V_{as}(\tau)\right] h_{\mu\nu}(t,\vec{x}) \mathrm{T}\exp\left[-i\int^t\ud{\tau} V_{as}(\tau)\right] \\
\nonumber
& = & \exp\left[i\int^t\ud{\tau} V_{as}(\tau)\right] h_{\mu\nu}(t,\vec{x}) \exp\left[-i\int^t\ud{\tau} V_{as}(\tau)\right] \\
\nonumber
& = & h_{\mu\nu} + i\int^t\ud{\tau}\int\ud[3]{y} [h_{\sigma\lambda}(\tau,\vec{y}),h_{\mu\nu}(t,\vec{x})] T_{as}^{\sigma\lambda}(\tau,\vec{y}) \\
& = & h_{\mu\nu} - \int^t\ud{\tau}\int\ud[3]{y} D(\tau-t,\vec{y}-\vec{x}) I_{\mu\nu;\sigma\lambda} T_{as}^{\sigma\lambda}(\tau,\vec{y}),
\end{IEEEeqnarray}
where $I_{\mu\nu;\sigma\lambda} = \eta_{\mu\sigma}\eta_{\nu\lambda} + \eta_{\mu\lambda}\eta_{\nu\sigma} - \eta_{\mu\nu}\eta_{\sigma\lambda}$ and
\begin{equation}
D(y-x) = D^{+}(y-x)-D^{+}(x-y),\quad D^{+}(y-x) = i\int\udpi[3]{k}\frac{1}{2k_0}e^{ik\cdot(y-x)}.
\end{equation}
In the second line we have used the fact that commutators of more than two factors of $h_{\mu\nu}$ vanish, along with the identity
\begin{equation} \label{CommutatorIdentity}
e^X Y e^{-X} = \sum_{n=0}^\infty \frac{1}{n!}\overbrace{[X,\dots[X}^n,Y]\dots].
\end{equation}
The object appearing in the simple form of the linearized Einstein equation is the trace-reversed perturbation $\bar{h}_{\mu\nu}$, so we compute
\begin{IEEEeqnarray}{rCl}
h_{as}(t,\vec{x}) & = & \eta^{\mu\nu}h^{as}_{\mu\nu}(t,\vec{x}) = h + 2\int^t\ud{\tau}\int\ud[3]{y} D(\tau-t,\vec{y}-\vec{x}) \eta_{\sigma\lambda} T_{as}^{\sigma\lambda}(\tau,\vec{y}) \\
\nonumber
\bar{h}^{as}_{\mu\nu}(t,\vec{x}) & = & h^{as}_{\mu\nu}(t,\vec{x}) - (1/2)h_{as}\eta_{\mu\nu}(t,\vec{x}) \\
\nonumber
& = & \bar{h}_{\mu\nu}(t,\vec{x}) - \int^t\ud{\tau}\int\ud[3]{y} D(\tau-t,\vec{y}-\vec{x}) [I_{\mu\nu;\sigma\lambda}-\eta_{\mu\nu}\eta_{\sigma\lambda}] T_{as}^{\sigma\lambda}(\tau,\vec{y}) \\
& = & \bar{h}_{\mu\nu}(t,\vec{x}) - 2 \int^t\ud{\tau}\int\ud[3]{y} D(\tau-t,\vec{y}-\vec{x})T^{as}_{\mu\nu}.
\end{IEEEeqnarray}
We remind the reader that the asymptotic fields in this section are all defined at large times.

The asymptotic field $\bar{h}^{as}_{\mu\nu}$ is seen to have the following properties:
\begin{enumerate}
\item
$\bar{h}^{as}_{\mu\nu}$ is sourced by $T^{as}_{\mu\nu}$:
\begin{equation}
\Box \bar{h}^{as}_{\mu\nu} = -2T^{as}_{\mu\nu} = -16\pi G T^{as}_{\mu\nu}.
\end{equation}
This follows because, for $\tau<t$, the function $D(\tau-t,\vec{y}-\vec{x})$ is equal to the retarded propogator and is therefore a Green's function for the wave operator.

\item 
Because $T^{as}_{\mu\nu}$ commutes with itself everywhere and with $h_{\mu\nu}$, the transformed operator $\bar{h}^{as}_{\mu\nu}$ has the same commutator as the free field $\bar{h}_{\mu\nu}$:
\begin{equation}
[\bar{h}^{as}_{\mu\nu}(x),\bar{h}^{as}_{\rho\sigma}(y)] = -iI_{\mu\nu;\rho\sigma}D(x-y).
\end{equation}
\end{enumerate}

The asymptotic gravitational field $\bar{h}^{as}_{\mu\nu}$ measures the asymptotic (i.e., classical) field sourced by a matter particle, in the sense that
\begin{equation}
[\bar{h}^{as}_{\mu\nu}(x),b^\dagger(q)] = 2G \frac{q_\mu q_\nu}{\sqrt{(q\cdot x)^2-q^2 x^2}}b^\dagger(q).
\label{AS}
\end{equation}
The coefficient of $b^{\dagger}$ on the RHS is just the classical linearized gravitational field of a point mass moving with momentum $\qvec$ \cite{Aichelburg-Sexl}. Thus, we see that with respect to the asymptotic operator, each particle is automatically created in association with its gravitational field.

Eq. \eqref{AS} follows straightforwardly by evaluating the commutators $[\rho(p),b^\dagger(q)]$
and $[T_{\mu\nu}^{as}(\tau,\vec{y}),b^\dagger(q)]$, which give
\begin{IEEEeqnarray}{rCl}
\nonumber
[\bar{h}_{\mu\nu}^{as}(t,\vec{x}),b^\dagger(q)] & = & -2\int^t\ud{\tau}\ud[3]{y} D(\tau-t,\vec{y}-\vec{x}) \delta^{(3)}(\vec{y}-\tau\vec{p}/p_0)\frac{q_\mu q_\nu}{2q_0} b^\dagger(q) \\
& = & \frac{2}{(2\pi)^3}\int^t\ud{\tau}\int\!\frac{\mathrm{d}^3 k}{2k_0} \sin\left(k\cdot\left(\frac{\tau}{q_0}q - x \right)\right) \frac{q_\mu q_\nu}{q_0} b^\dagger(q).
\end{IEEEeqnarray}
This integral is computed in the appendix of \cite{FK}, with the result
\begin{IEEEeqnarray}{rCl}
\nonumber
\int^t\ud{\tau}\int\!\frac{\mathrm{d}^3 k}{2k_0} \sin\left(k\cdot\left(\frac{\tau}{q_0}q - x \right)\right) & = & \int^t\ud{\tau} (2\pi^2) \delta\left(\left(\frac{\tau}{q_0}q-x\right)^2\right) \\
& = & (2\pi^2) \frac{q_0}{2\sqrt{(q\cdot x)^2 - q^2 x^2}}.
\end{IEEEeqnarray}
Substituting this and combining factors of $2\pi$, etc., we have
\begin{IEEEeqnarray}{rCl}
\nonumber
[\bar{h}^{as}_{\mu\nu}(x),b^\dagger(q)] & = & \frac{1}{4\pi}\frac{q_\mu q_\nu}{\sqrt{(q\cdot x)^2-q^2 x^2}}b^\dagger(q) \\
& = & 2G \frac{q_\mu q_\nu}{\sqrt{(q\cdot x)^2-q^2 x^2}}b^\dagger(q),
\end{IEEEeqnarray}
upon restoring the Newton constant.

Finally, consider the free (scalar) matter field
\begin{equation}
\phi(x) = \int\!\frac{\mathrm{d}^3 p}{(2\pi)^{3/2}}\frac{1}{\sqrt{2p_0}}(b(p)e^{ipx}+b^\dagger(p)e^{-ipx}).
\end{equation}
As before, we wish to compute the asymptotic field $\phi_{as}(t,\vec{x}) = Z^\dagger(t)\phi(t,\vec{x})Z(t)$.  Note that the commutator term $e^{i\Phi}$ does not commute with $\phi(t,\vec{x})$ as it did with $h_{\mu\nu}(t,\vec{x})$, since we have the nonzero commutator
\begin{equation}
[:\rho(p)\rho(q):,b^\dagger(r)] = -\delta^{(3)}(\vec{r}-\vec{p})b^\dagger(r)\rho(q) -\delta^{(3)}(\vec{r}-\vec{q})b^\dagger(r)\rho(p)
\end{equation}
and so
\begin{equation}
[-i\Phi(t),b^\dagger(r)] = b^\dagger(r) \left[ \frac{i}{32\pi}I(t)\int\ud[3]{p}\frac{[2(p\cdot r)^2-m^4]}{\sqrt{(p\cdot r)^2-m^4}}\rho(p) \right]
\end{equation}
after relabeling the integration variable $q\rightarrow p$ in the first term, where $I(t)$ is the integral defined in eq. \eqref{tauintegral}.  Then since $\Phi$ commutes with $\rho(p)$, successive applications of $\Phi$ just produce more factors:
\begin{equation}
\overbrace{[-i\Phi(t),\dots[-i\Phi(t),}^{n}b^\dagger(r)]\dots] =  b^\dagger(r) \left[ \frac{i}{32\pi}I(t)\int\ud[3]{p}\frac{[2(p\cdot r)^2-m^4]}{\sqrt{(p\cdot r)^2-m^4}}\rho(p) \right]^n
\end{equation}
and so, using the identity \eqref{CommutatorIdentity},
\begin{equation}
e^{-i\Phi(t)}b^\dagger(r) e^{i\Phi(t)} = b^\dagger(r) \exp\left[ \frac{i}{32\pi}I(t)\int\ud[3]{p}\frac{[2(p\cdot r)^2-m^4]}{\sqrt{(p\cdot r)^2-m^4}}\rho(p) \right]
\end{equation}
and similarly
\begin{equation}
e^{-i\Phi(t)}b(r) e^{i\Phi(t)} = \exp\left[ \frac{-i}{32\pi}I(t)\int\ud[3]{p}\frac{[2(p\cdot r)^2-m^4]}{\sqrt{(p\cdot r)^2-m^4}}\rho(p) \right] b(r).
\end{equation}

There is also a contribution from the factor $e^{R(t)}$, since
\begin{equation}
[V_{as}(\tau),b^\dagger(r)] = -h^{\mu\nu}(\tau r/r_0)\frac{r_\mu r_\nu}{2r_0} b^\dagger(r).
\end{equation}
Since $V_{as}$ commutes with $\rho(\pvec)$, this contribution is independent of the previous one, and it exponentiates similarly:
\begin{IEEEeqnarray}{rCl}
e^{-R(t)} b^\dagger(r) e^{R(t)} & = & \exp\left[-i\int\ud{\tau}h^{\mu\nu}(\tau r/r_0)\frac{r_\mu r_\nu}{2r_0}\right]b^\dagger(r), \\
e^{-R(t)} b(r) e^{R(t)} & = & \exp\left[+i\int\ud{\tau}h^{\mu\nu}(\tau r/r_0)\frac{r_\mu r_\nu}{2r_0}\right]b(r).
\end{IEEEeqnarray}
Altogether, we have
\begin{multline}
\phi_{as}(t,\vec{x}) = Z^\dagger(t) \phi(t,\vec{x}) Z(t) = \int\frac{\mathrm{d}^3 r}{(2\pi)^{3/2}}\frac{1}{\sqrt{2r_0}}\Bigg\{ \exp\left[i\int\ud{\tau}h^{\mu\nu}(\tau r/r_0)\frac{r_\mu r_\nu}{2r_0}\right]  \\ \times \exp\left[ \frac{-i}{32\pi}I(t)\int\ud[3]{p}\frac{[2(p\cdot r)^2-m^4]}{\sqrt{(p\cdot r)^2-m^4}}\rho(p) \right] b(r) e^{irx} + \text{h.c.}\Bigg\},
\end{multline}
which contains contributions from the phase operator and from the soft component of the gravitational field. Due to these contributions, the asymptotic field associated with the scalar particle does not create or annihilate single particle states, in agreement with the result Eq. (\ref{AS}).

\section{Coherent states: properties and constructions}
\label{sec:coherent}
In this section, we define the IR-finite S matrix for perturbative quantum gravity and the associated space of asymptotic coherent states. We demonstate the desirable Lorentz and gauge transformation properties of that space.  Finally, we construct, following \cite{Chung}, the states that will be used in the next section to show the cancellation of infrared divergences at the level of S matrix elements in the Faddeev-Kulish framework. 

Using the operator of asymptotic dynamics $U_{as}(t) = e^{-iH_0 t}Z(t)$ in place of the free time evolution operator, we define the asymptotic S matrix
\begin{equation}
S_A = \lim_{t\rightarrow\infty} Z^{\dagger}(t)S_D Z(t),
\label{smatrix}
\end{equation}
where $S_D$ is the standard Dyson S matrix.

Note that matrix elements of Eq. \eqref{smatrix} may be viewed in two equivalent ways: as the matrix elements of the asymptotic S matrix $S_A$ between standard Fock states, or as the matrix elements of the standard Dyson S matrix between the states of the asymptotic space $\Hilbert_{as}=Z(t)\Hilbert_{Fock}$. This is the space of the coherent states of the theory.\footnote{Note that this differs from the definition proposed in \cite{FK}, which uses $\Hilbert_{as}=e^{-R(t)}\Hilbert_F$. For a discussion of the rationale behind this difference, see Appendix A of \cite{Contopanagos}.  In particular, the states used by Chung \cite{Chung} to show the cancellation of divergences in QED belong to the space $e^{+R(t)}\Hilbert_F$, not $e^{-R(t)}\Hilbert_F$.} In the following subsections we will show that it is Lorentz- and gauge-invariant.  First, we introduce a useful alternative characterization of $\Hilbert_{as}$.

Recall that the asymptotic operator $Z(t)=e^{i\Phi(t)}e^{R(t)}$ is given for large $t$ by 
\begin{equation}
R(t) = \frac{1}{(2\pi)^{3/2}}\int\!\frac{\mathrm{d}^3k \mathrm{d}^3p}{2\sqrt{2k^0}} \frac{p^\mu p^\nu}{p\cdot k} \left(a_{\mu\nu}(k)e^{i\frac{k\cdot p}{p^0}t} - \text{h.c.}\right)\rho(p)
\end{equation}
and, taking the limit $t\rightarrow\infty$ in dimensional regularization,
\begin{equation}
\Phi(\infty)=-\frac{1}{16\pi}\int\ud[3]{p}\ud[3]{q} :\rho(p)\rho(q): \frac{2(p\cdot q)^2-m^4}{\sqrt{(p\cdot q)^2-m^4}} \frac{1}{\epsilon}.
\end{equation}
Although formally $Z^\dagger(t) = Z^{-1}(t)$, $Z(t)$ does not define a unitary transformation in the Fock space because the operator $e^{R(t)}$ creates unbounded numbers of low-energy gravitons. It is this low-energy behavior that distinguishes the space $\Hilbert_{as}$, rather than the precise operator $R(t)$. Consider instead an operator $R_f$ of form similar to $R(t)$, but characterized by an IR function $f^{\mu\nu}(k,p)$:
\begin{equation}
R_f = \frac{1}{(2\pi)^{3/2}}\int\ud[3]{p}\!\frac{\mathrm{d}^3k}{2\sqrt{2k_0}}(f^{\mu\nu}(k,p)a^\dagger_{\mu\nu}(k) - \text{h.c.})\rho(p).
\end{equation}
Using the Baker-Campbell-Hausdorff (BCH) formula, we can write
\begin{equation}
e^{R_f} = e^{R(t)}e^{R_f-R(t)}e^{(1/2)[R_f,R(t)]}.
\label{BCH}
\end{equation}
If the last two factors are well-defined and unitary within the Fock space, then $R_f$ provides an alternative definition of the space of asymptotic states:
\begin{IEEEeqnarray}{rCl}
\nonumber
e^{R_f}\Hilbert_F & = & e^{R(t)}e^{R_f-R(t)}e^{(1/2)[R_f,R(t)]}\Hilbert_F \\
\nonumber
& = & e^{R(t)}\Hilbert_F \\
& = & \Hilbert_{as}.
\end{IEEEeqnarray}
The freedom to choose (subject to constraints) the IR function $f^{\mu\nu}(k,p)$ will become useful in defining the Chung states in subsection \ref{sec:Chung}.

The unitarity of the factors appearing in eq. \eqref{BCH} imposes convergence constraints on the functions $f^{\mu\nu}(k,p)$ that can be used to characterize $\Hilbert_{as}$. The commutator factor involves the integral
\begin{equation}
\int\!\frac{\mathrm{d}^3k}{2k^0}\left( f^*_{\mu\nu}(k,p)\frac{q_\mu q_\nu}{k\cdot q}e^{-i\frac{k\cdot q}{q^0}t} - \frac{1}{2}f^{*\mu}_\mu(k,p)\frac{q^\mu q_\mu}{k\cdot q}e^{-i\frac{k\cdot q}{q^0}t} -\text{h.c.} \right),
\label{condition1}
\end{equation}
for the convergence of which (in the IR) it suffices to take $f_{\mu\nu}(k,p)$ real.  To check the unitarity of the factor $e^{R_f-R(t)}$, we can use BCH again to normal order.  The commutator factor here involves
\begin{equation}
\int\!\frac{\mathrm{d}^3k}{2k^0}\left\{\left|f_{\mu\nu}(k,p) - \frac{p_\mu p_\nu}{k\cdot p}e^{-i\frac{k\cdot p}{p^0}t} \right|^2 - \frac{1}{2}\left|f^\mu_\mu(k,p) - \frac{p^2}{k\cdot p}e^{-i\frac{k\cdot p}{p^0}t} \right|^2 \right\}.
\label{condition2}
\end{equation}
The convergence of the integral \eqref{condition2} will determine the low-energy behavior of $f_{\mu\nu}$.

We now use the ideas developed here to show that $\Hilbert_{as}$ is Lorentz- and gauge-invariant and that a subspace of physical states may be defined within it by the usual Gupta-Bleuler method.

\subsection{Gauge invariance}
\label{sec:gauge}
A gauge transformation takes
\begin{equation}
a_{\mu\nu}(k) \rightarrow a_{\mu\nu}(k) + k_\mu \eta_\nu + k_\nu \eta_\mu
\end{equation}
for some vector $\eta$.  The phase operator $\Phi(\infty)$ is unaffacted, but the transformation takes $W=e^{R(t)}$ to
\begin{IEEEeqnarray}{rCl}
\nonumber
W' & = & \exp\left\{\frac{1}{(2\pi)^{3/2}}\int\!\frac{\mathrm{d}^3k \mathrm{d}^3p}{2\sqrt{2k^0}} \frac{p^\mu p^\nu}{p\cdot k} \left((a_{\mu\nu}(k)+ k_\mu \eta_\nu + k_\nu \eta_\mu)e^{i\frac{k\cdot p}{p^0}t} - \text{h.c.}\right)\rho(p)\right\} \\
& = & W \times \exp\left\{\frac{1}{(2\pi)^{3/2}}\int\!\frac{\mathrm{d}^3k \mathrm{d}^3p}{\sqrt{2k^0}}\left((p\cdot\eta) e^{i\frac{k\cdot p}{p^0}t} - \text{h.c.}\right)\rho(p)\right\}.
\end{IEEEeqnarray}

Since the extra factor is unitary in the Fock space, the space of asymptotic states is invariant.

\subsection{Lorentz invariance}
\label{sec:lorentz}
Now consider the behavior of $Z(t)$ under a Lorentz transformation. The phase operator is again of no concern; its action on Fock space states just produces the manifestly Lorentz-invariant phase factor
\begin{equation}
-\frac{i}{16\pi}\frac{2(p_n\cdot p_m)^2-m^4}{\sqrt{(p_n\cdot p_m)^2-m^4}} \frac{1}{\epsilon}
\end{equation}
for each pair of particles.  To compute the Lorentz transformation properties of $R(t)$, recall that a unitary representation $U(\Lambda)$ of the Lorentz group is an active transformation; it acts only on states and operators and (by linearity) not on c-number functions, even if they explicitly involve coordinates or components of momenta.  So we only need to know how the creation and annihilation operators for gravitons and scalars transform:
\begin{IEEEeqnarray}{rCl}
U(a,\Lambda)a_{\mu\nu}(k)U^\dagger(a,\Lambda) & = & \sqrt{\frac{(\Lambda k)_0}{k_0}}(\Lambda^{-1})_\mu^{\phantom{\mu}\rho}(\Lambda^{-1})_\nu^{\phantom{\nu}\sigma}a_{\rho\sigma}(\Lambda k)e^{-2i\Lambda k\cdot a}, \\
U(a,\Lambda)\rho(p)U^\dagger(a,\Lambda) & = & \frac{(\Lambda p)_0}{p_0}\rho(\Lambda p).
\end{IEEEeqnarray}
The factors of $p_0$ or $k_0$ result from the normalization of one-particle states; see \cite{WeinbergBook}.  The phase shift of the graviton operator is twice that of the photon because it is proportional to the helicity.  

Applying the Lorentz transformation to $W$,
\begin{equation}
W = \exp\left\{\frac{1}{2(2\pi)^{3/2}}\int\!\frac{\mathrm{d}^3k}{2k_0}\frac{\mathrm{d}^3p}{2p_0} (2p_0)\sqrt{2k_0} \frac{p^\mu p^\nu}{p\cdot k} \left(a_{\mu\nu}(k)e^{i\frac{k\cdot p}{p^0}t} - \text{h.c.}\right)\rho(p) \right\},
\end{equation}
\begin{multline}
U(a,\Lambda)WU^\dagger(a,\Lambda) = \exp\bigg\{\frac{1}{2(2\pi)^{3/2}}\int\!\frac{\mathrm{d}^3k}{2k_0}\frac{\mathrm{d}^3p}{2p_0} \left(2(\Lambda p)_0\right)\sqrt{2(\Lambda k)_0} \frac{p^\mu p^\nu}{p\cdot k} \\ 
\times (\Lambda^{-1})_\mu^{\phantom{\mu}\rho}(\Lambda^{-1})_\nu^{\phantom{\nu}\sigma} \left(a_{\rho\sigma}(\Lambda k)e^{i\frac{k\cdot p}{p^0}t -2i\Lambda k\cdot a} - \text{h.c.}\right)\rho(\Lambda p) \bigg\}.
\end{multline}
We can associate the Lorentz transformation matrices with the preceding momenta $p^\mu p^\nu$, and write the Lorentz-invariant measures and scalar products in a transformed form:
\begin{multline}
U(a,\Lambda)WU^\dagger(a,\Lambda) = \exp\bigg\{\frac{1}{2(2\pi)^{3/2}}\int\!\frac{\mathrm{d}^3(\Lambda k)}{2(\Lambda k)_0}\frac{\mathrm{d}^3(\Lambda p)}{2(\Lambda p)_0} \left(2(\Lambda p)_0\right)\sqrt{2(\Lambda k)_0} \frac{(\Lambda p)^\rho (\Lambda p)^\sigma}{(\Lambda p)\cdot (\Lambda k)} \\ 
\times \left(a_{\rho\sigma}(\Lambda k)e^{i\frac{(\Lambda k)\cdot (\Lambda p)}{p^0}t -2i\Lambda k\cdot a} - \text{h.c.}\right)b^\dagger(\Lambda p)b(\Lambda p) \bigg\}.
\end{multline}
Finally, rename $p\rightarrow\Lambda^{-1}p$, $k\rightarrow\Lambda^{-1}k$:
\begin{multline}
U(a,\Lambda)WU^\dagger(a,\Lambda) = \exp\bigg\{\frac{1}{2(2\pi)^{3/2}}\int\!\frac{\mathrm{d}^3k}{2k_0}\frac{\mathrm{d}^3p}{2p_0} \left(2p_0\right)\sqrt{2k_0} \frac{p^\rho p^\sigma}{p\cdot k} \\ 
\times \left(a_{\rho\sigma}(k)e^{i\frac{k\cdot p}{(\Lambda^{-1}p)^0}t -2ik\cdot a} - \text{h.c.}\right)b^\dagger(p)b(p) \bigg\}.
\end{multline}
This is an operator of the form $e^{R_f}$, with the same IR behavior as W, so it generates the same space of asymptotic states. This establishes that $\Hilbert_{as}$ is Lorentz invariant.

\subsection{Construction of physical (Chung) states}
\label{sec:Chung}
In order to show the cancellation of IR divergences, we adopt the point of view that the Dyson S matrix acts in the space of coherent states and follow Chung's approach for QED in \cite{Chung}. The coherent state used by Chung belongs to the space $\Hilbert'_{as}$, but it is not the state $e^{R(t)}b^\dagger(p)|0\rangle$.  Instead, Chung uses a physical state gauge-equivalent to one of the form $e^{R_f}b^\dagger(p)|0\rangle$. In this subsection we will construct such states for perturbative quantum gravity.

First, we recall that in the Fock space, the physical states satisfy the condition \cite{Gupta}
\begin{equation}
(k^\mu a_{\mu\nu}(k) - (1/2)k_\nu a^\rho_\rho(k))|\Psi\rangle = 0.
\label{physicalcondition}
\end{equation}
We define the physical subspace of $\Hilbert_{as}$ by the same condition.  We would like to be able to obtain the physical asymptotic states by a transformation of the physical Fock states, which will be possible if we can find an operator of the form $e^{R_f}$ that commutes with \eqref{physicalcondition}.  This requires that
\begin{IEEEeqnarray}{rCl}
\nonumber 
0 & = & [k^\mu a_{\mu\nu} - (1/2)k_\nu \eta^{\rho\sigma}a_{\rho\sigma}, f^{\alpha\beta}a^\dagger_{\alpha\beta}] \\
\nonumber
& = & k^\mu f_{\mu\nu} + k^\mu f_{\nu\mu} - k_\nu f^\alpha_\alpha - (1/2)k_\nu(f^\alpha_\alpha+f^\alpha_\alpha - \eta^{\rho\sigma}\eta_{\rho\sigma}f^\alpha_\alpha) \\
& = & 2 k^\mu f_{\mu\nu}
\end{IEEEeqnarray}
for $f_{\mu\nu}$ symmetric.  

We also need $f$ to satisfy the convergence constraint \eqref{condition2}. This requires $f$ to have a singularity that will cancel against $p_\mu p_\nu / kp$, so put
\begin{equation}
f_{\mu\nu}(k,p) = \left[\frac{p_\mu p_\nu}{k\cdot p} + c_{\mu\nu}\right]\phi(k,p),
\end{equation}
where $k^\mu c_{\mu\nu} = -p_\nu$ and where $\phi$ is some smoothing function with $\phi=1$ for small $k$.  To avoid producing additional singular terms in the integral, we need $c_{\mu\nu}c^{\mu\nu} - (1/2) (c^\mu_\mu)^2 = 0$.  Later, we will modify $f_{\mu\nu}$ to make it a linear combination of graviton polarization tensors $\epsilon^{n}_{\mu\nu}(k)$; at that point, it will also be useful if $c_{\mu\nu}\epsilon^{n\mu\nu}(k)=0$.  Make the ansatz
\begin{IEEEeqnarray}{rCl}
\nonumber
c_{00} & = & x_0, \\
\nonumber
c_{0i} & = & x_1p_i + x_2k_i, \\
c_{ij} & = & x_3p_ip_j + x_4 k_ik_j,
\end{IEEEeqnarray}
with the coefficients $x$ to be determined.  Since the graviton polarization tensors are purely spatial and orthogonal to $k$, $c_{\mu\nu}$ will be orthogonal to polarization tensors if we choose $x_3=0$.  Then $k^\mu c_{\mu\nu}=0$ requires (using the on-shell condition $k_ik_i = \kvec^2 = k_0^2$)
\begin{IEEEeqnarray}{rCl}
\nonumber
-p_0 & = & k_0x_0-x_1\kvec\cdot\pvec - k_0^2x_2, \\
-p_i & = & k_0x_1p_i + k_0x_2k_i - x_4 k_0^2k_i,
\end{IEEEeqnarray}
so
\begin{IEEEeqnarray}{rCl}
\nonumber
x_1 & = & -\frac{1}{k_0}, \\
\nonumber
x_4 & = & \frac{x_2}{k_0}, \\
\nonumber
x_0 & = & -\frac{p_0}{k_0} + x_1\frac{\pvec\cdot\kvec}{k_0} + x_2k_0 \\
    & = & -\frac{p_0}{k_0} - \frac{\pvec\cdot\kvec}{k_0^2} + x_2k_0.
\end{IEEEeqnarray}
This leaves one remaining coefficient, $x_2$, and one constraint, $c_{\mu\nu}c^{\mu\nu} - (1/2) (c^\mu_\mu)^2 = 0$.  We now have
\begin{IEEEeqnarray}{rCl}
\nonumber
c_{00} & = & -\frac{p_0}{k_0}, \\
\nonumber
c_{0i} & = & -\frac{p_i}{k_0} + x_2k_i, \\
c_{ij} & = & x_2\frac{k_ik_j}{k_0}.
\end{IEEEeqnarray}
It looks as if the remaining constraint will be quadratic in $x_2$ and so might not always have a solution, but in fact the quadratic terms cancel, leaving $x_2$ (an algebraic mess but) determined.

In the next section, we will consider one-scalar gravitational potential scattering in detail in order to demonstrate the cancellation of divergences. In order to write a one-scalar asymptotic state $|\Psi\rangle = e^{R_f}b^{\dagger}(p)|0\rangle$ in the Chung form, we first need to arrange for $f_{\mu\nu}(k,p)$ to be a linear combination of graviton polarization tensors.  This can be achieved using the residual freedom
\begin{equation}
f_{\mu\nu} \rightarrow f'_{\mu\nu} = f_{\mu\nu} + k_\mu\lambda_\nu + k_\nu\lambda_\mu - (k\cdot\lambda)\eta_{\mu\nu},
\end{equation}
which preserves the physical state condition.  It is a standard result in linearized GR that a symmetric, transverse metric perturbation $h_{\mu\nu}(k)$ can be made traceless and purely spatial $h_{\mu 0}=0$ using transformations of this kind.  But we first need to show that such a transformation, which may disturb the conditions imposed above in the determination of $c_{\mu\nu}$, nonetheless keeps our state in the space of asymptotic states.

First use Baker-Campbell-Hausdorff to write $|\Psi\rangle_{as}$ in normal-ordered form:
\begin{multline}
|\Psi\rangle = \exp\left\{-\frac{1}{2(2\pi)^3}\int\!\frac{\mathrm{d}^3k}{8k_0}\big(2f^{\mu\nu}f_{\mu\nu} - |f^\mu_\mu|^2\big)\right\} \\ \times \exp\left\{\frac{1}{(2\pi)^{3/2}}\int\!\frac{\mathrm{d}^3k}{2\sqrt{2k_0}}f^{\mu\nu}a^\dagger_{\mu\nu}\right\} \exp\left\{-\frac{1}{(2\pi)^{3/2}}\int\!\frac{\mathrm{d}^3k}{2\sqrt{2k_0}}f^{\mu\nu}a_{\mu\nu}\right\} b^{\dagger}(p)|0\rangle.
\end{multline}
Now consider the effect of the tranformation on each of the three factors.  The commutator factor is invariant,
\begin{IEEEeqnarray}{rCl}
\nonumber
2f^{\mu\nu}f_{\mu\nu} - (f^\mu_\mu)^2 & \rightarrow & 2\big(f^{\mu\nu}f_{\mu\nu} - 2(k\cdot\lambda)f^\mu_\mu + 2(k\cdot\lambda)^2\big) - \big((f^\mu_\mu)^2 + 4(k\cdot\lambda)f^\mu_\mu - 4(k\cdot\lambda)^2\big) \\
& = & 2f^{\mu\nu}f_{\mu\nu} - (f^\mu_\mu)^2,
\end{IEEEeqnarray}
provided that $k^\mu f_{\mu\nu}=0$, as in our case.  The factor containing creation operators picks up a multiplicative factor
\begin{equation}
\exp\left\{\frac{1}{(2\pi)^{3/2}}\int\!\frac{\mathrm{d}^3k}{2\sqrt{2k_0}} 2\lambda^\nu\left(k^\mu a^\dagger_{\mu\nu}(k) - \frac{1}{2}k_\nu a^{\rho\dagger}_{\rho}(k)\right)\right\}.
\end{equation}
Since the expression in parentheses above is conjugate to the physical state condition, terms beyond zeroth order in the exponential will be orthogonal to all physical states, i.e. they will be spurious.  But in Gupta-Bleuler quantization we identify states that differ by $|\chi\rangle$ if $|\chi\rangle$ is spurious.  So this factor has no effect.  Finally, the annihilation part picks up a similar factor
\begin{equation}
\exp\left\{-\frac{1}{(2\pi)^{3/2}}\int\!\frac{\mathrm{d}^3k}{2\sqrt{2k_0}} 2\lambda^\nu\left(k^\mu a_{\mu\nu}(k) - \frac{1}{2}k_\nu a^{\rho}_{\rho}(k)\right)\right\},
\end{equation}
in which the terms beyond the zeroth order vanish on physical states.  This shows that we are free to make the residual transformation, and so we can choose $f'_{\mu\nu}$ to be traceless and spatial.

We can now write $f'_{\mu\nu}$ as a linear combination of graviton polarization tensors:
\begin{equation}
f'_{\mu\nu}(k,p) = F^1(k,p)\epsilon^1_{\mu\nu}(k) + F^2(k,p)\epsilon^2_{\mu\nu}(k).
\end{equation}
Normalizing the polarization tensors $\epsilon^n_{\mu\nu}\epsilon^{n\mu\nu}=1$ and recalling that 
\begin{equation}
k^\mu\epsilon_{\mu\nu} = c^{\mu\nu}\epsilon_{\mu\nu} = \eta^{\mu\nu}\epsilon_{\mu\nu} = 0,
\end{equation}
we can extract
\begin{equation}
F^n = \frac{p^\mu p^\nu}{p\cdot k}\epsilon^n_{\mu\nu}\phi(k,p).
\end{equation}
 
 \section{Cancellation of infrared divergences in gravitational potential scattering}
 \label{sec:cancellation}

\begin{figure}
\begin{center}
\includegraphics[width=0.25\textwidth]{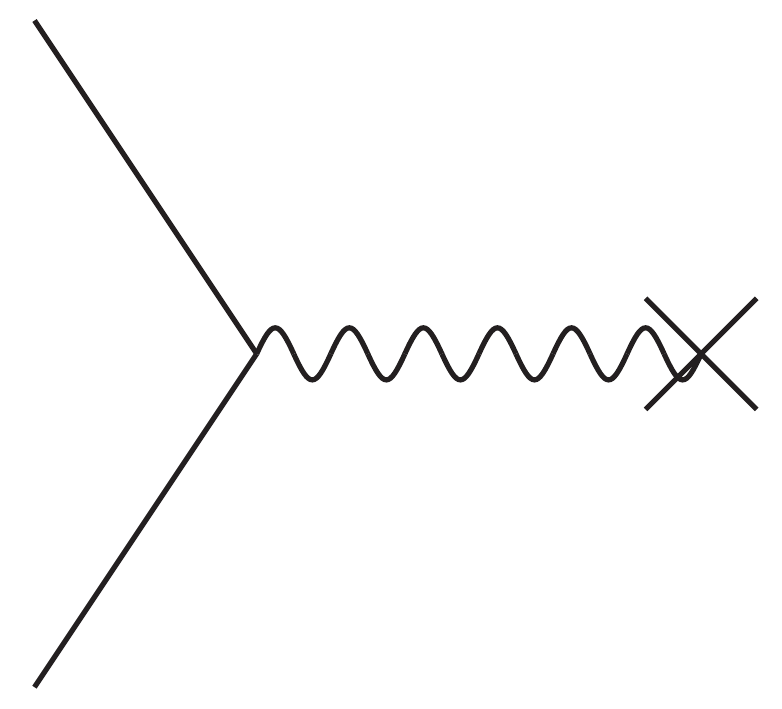}
\end{center}
\caption{Gravitational potential scattering at tree level.}
\label{process}
\end{figure}

To illustrate the cancellation of IR divergences in matrix elements of the coherent states defined above, we consider the simple case of gravitational potential scattering, shown at tree level in figure \ref{process}.  Our work follows the treatment of Chung in QED \cite{Chung}.  In section \ref{sec:oneloop} we consider divergences at order $\kappa^2$, and in \ref{sec:allorders} we show that the cancellation extends to all orders.

	\subsection{Cancellation at one loop}
	\label{sec:oneloop}

The asymptotic state is 
\begin{equation}
\Psi = e^{+R_f'}b^{\dagger}(p)|0\rangle = \exp\left[\frac{1}{(2\pi)^{3/2}}\int\!\frac{\mathrm{d}^3 k}{2\sqrt{2k_0}}(f'_{\mu\nu}(k,p)a^{\dagger\mu\nu}(k) - \mathrm{h.c.})\right]b^\dagger(p)|0\rangle,
\end{equation}
or, using Baker-Campbell-Hausdorff to normal order and omitting a factor involving graviton annihilation operators,
\begin{IEEEeqnarray}{rCl}
\nonumber
\Psi & = & \exp\left[-\frac{1}{2}\frac{1}{(2\pi)^3}\int\!\frac{\mathrm{d}^3 k}{8k_0} \big(2|f'_{\mu\nu}|^2 - |f'^\mu_\mu|^2 \big) \right] \exp\left[\frac{1}{(2\pi)^{3/2}}\int\!\frac{\mathrm{d}^3 k}{2\sqrt{2k^0}} f'^{\mu\nu}a^\dagger_{\mu\nu} \right] b^\dagger(p) |0\rangle \\
\nonumber
& = & \exp\left[-\frac{1}{(2\pi)^3}\int\!\frac{\mathrm{d}^3 k}{8k_0}\sum_n |F^n|^2\right] \exp\left[\frac{1}{(2\pi)^{3/2}}\int\!\frac{\mathrm{d}^3 k}{2\sqrt{2k_0}} \sum_n F^n \epsilon^n_{\mu\nu} a^{\dagger\mu\nu}(k)\right] b^\dagger(p)|0\rangle. \\
\end{IEEEeqnarray}
Using a similar notation to Chung, we define
\begin{equation}
S_i^n = \frac{1}{2\sqrt{(2\pi)^3 2k_0}}F^n(p_i)
\end{equation}
so that the initial state is, to lowest order,
\begin{equation}
|i\rangle = \left(1-\int\ud[3]{k}\sum_n|S_i^n|^2\right)\left(1+\int\ud[3]{k}\sum_n S_i^n \epsilon^n_{\mu\nu}a^{\dagger\mu\nu}(k)\right)b^\dagger(p_i)|0\rangle,
\end{equation}
and similarly for the final state.

\begin{figure}
\begin{center}
\includegraphics[width=\textwidth]{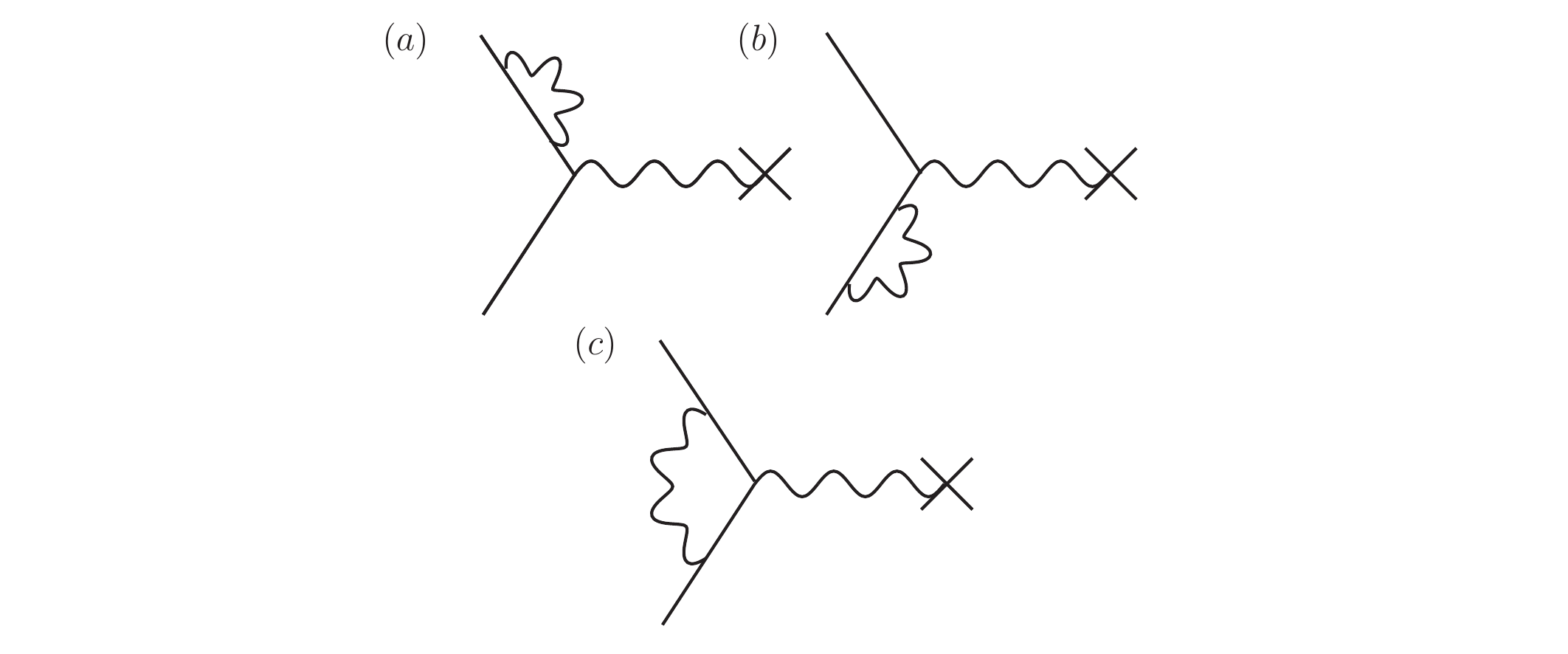}
\end{center}
\caption{Diagrams at order $\kappa^2$ with graviton loops.}
\label{loop}
\end{figure}

\begin{figure}
\begin{center}
\includegraphics[width=\textwidth]{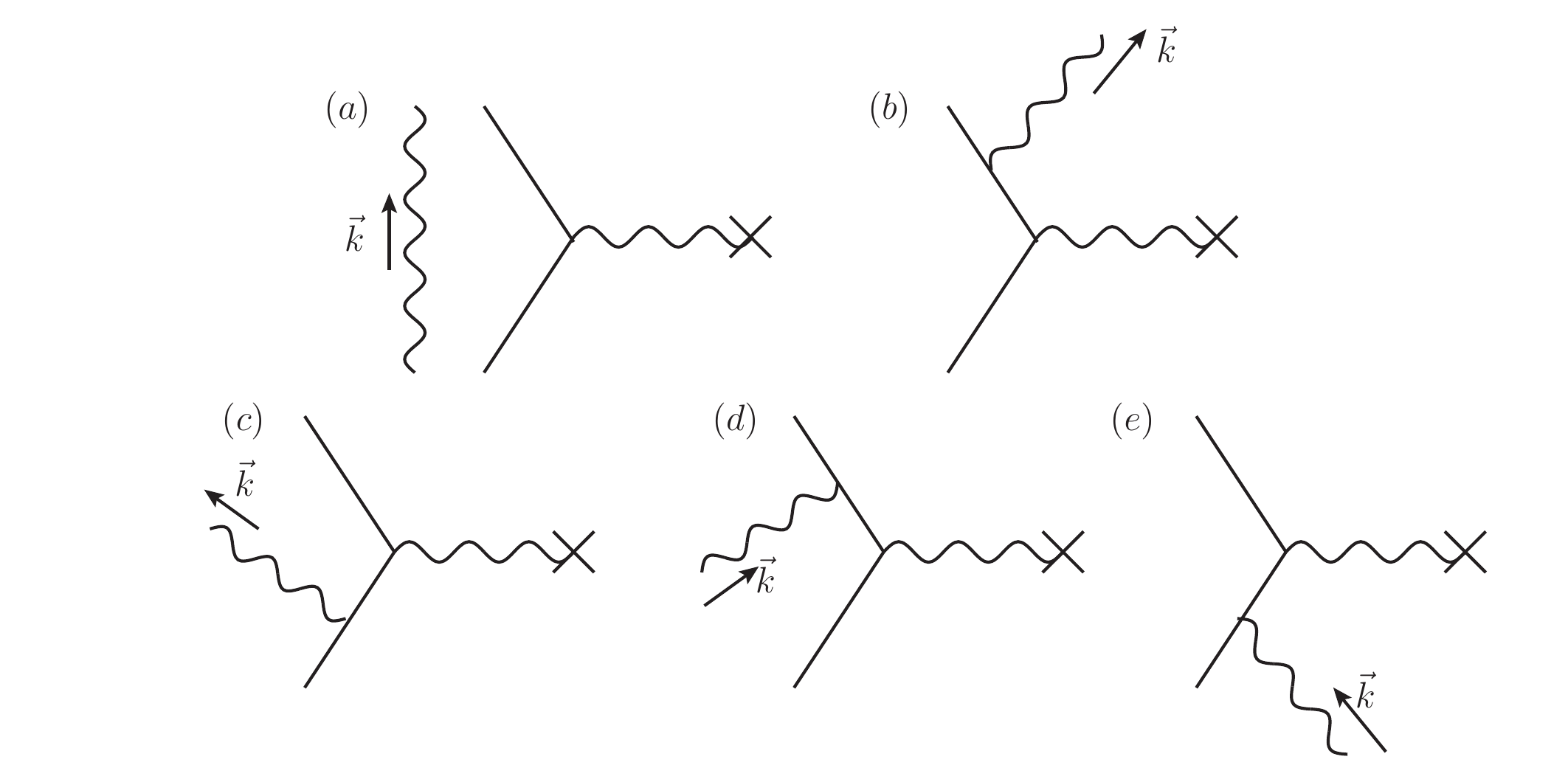}
\end{center}
\caption{Diagrams at order $\kappa^2$ with external soft gravitons.}
\label{real}
\end{figure}

Consider the matrix element for this state to scatter from a gravitational potential.  The following contributions are up by $\kappa^2$ relative to the tree-level process with no external soft gravitons:
\begin{enumerate}
\item Diagrams containing one soft graviton loop.  These diagrams correspond directly to those in QED and are computed by Weinberg in \cite{Weinberg}.  They are shown in figure \ref{loop}.
\item The tree diagram, with no external gravitons, but with the second-order term in the normalization of either the initial or the final state.
\item The tree diagram with an additional disconnected soft graviton line.  This is up by $\kappa$ in both the initial and the final state, and so to second order it can contain no extra vertices.  It corresponds to diagram (a) of figure \ref{real}. At the end of this subsection we will comment briefly on the need to include such contributions.
\item Tree-level diagrams with one external soft graviton in either the initial or the final state.  The asymptotic state contributes one factor of $\kappa$ and the additional vertex contributes another factor.  These correspond to the diagrams (b)-(e) of figure \ref{real}.
\end{enumerate}

We now compute these contributions in turn.
\begin{enumerate}
\item
Weinberg \cite{Weinberg} computes the virtual soft graviton corrections. Although he uses only linearized interactions, the results of \cite{Coll} indicate that his results are exact as far as IR divergences are concerned, since quadratic and higher vertices do not contribute.  At one loop, then, virtual soft gravitons contribute a factor $(1/2)\int\ud[4]{k}B(k)$, where
\begin{equation}
B(k) = \frac{1}{2(2\pi)^4}\sum_{n,m}\eta_n\eta_m[(p_n\cdot p_m)^2-(1/2)m_n^2 m_m^2] J_{mn}(k),
\end{equation}
the sum runs over initial $(\eta=-1)$ and final $(\eta=+1)$ state particles, and
\begin{IEEEeqnarray}{rCl}
\nonumber
\real\int\ud[4]{k}J_{mn}(k) & = & \real\int\ud[4]{k}\frac{i}{(k^2-i\epsilon)(p_n\cdot k -i\eta_n\epsilon)(p_m\cdot k +i\eta_m\epsilon)} \\
& = & -\pi\int\!\frac{\mathrm{d}^3k}{k^0}\frac{1}{(p_n\cdot k)}\frac{1}{(p_m\cdot k)}.
\end{IEEEeqnarray}
The imaginary part is canceled by the asymptotic phase operator.  The real part is
\begin{equation}
\frac{1}{2}\int\ud[4]{k}\left[\real B(k)\right] = -\frac{1}{64\pi^3}\sum_{n,m}\int\!\frac{\mathrm{d}^3k}{k^0}\frac{\eta_n\eta_m[(p_n\cdot p_m)^2-(1/2)m_n^2 m_m^2]}{(p_n\cdot k)(p_m\cdot k)}.
\end{equation}
In the case at hand, with one initial-state particle ($\eta=-1$, momentum $p_i$) and one final state particle ($\eta=1$, momentum $p_f$), we have
\begin{equation}
\label{virtualB}
-\frac{1}{64\pi^3}\int\!\frac{\mathrm{d}^3k}{k^0}\left[\frac{p_f^4 - (1/2)p_f^4}{(p_f\cdot k)^2} + \frac{p_i^4 - (1/2)p_i^4}{(p_i\cdot k)^2} - 2\left(\frac{(p_f \cdot p_i)^2 - (1/2)(p_f)^2(p_i)^2}{(p_f\cdot k)(p_\cdot k)}\right)\right].
\end{equation}

\item
The second-order terms in the normalization contribute a factor
\begin{equation}
-\sum_n \int\ud[3]{k}\left[|S_i^n|^2+|S_f^n|^2\right].
\end{equation}

\item
The disconnected graviton line gives a factor
\begin{IEEEeqnarray}{rCl}
\nonumber
& & \sum_{n,n'} \int\ud[3]{k}\ud[3]{k'}  S^n_i \epsilon^n_{\mu\nu} S_f^{n'} \epsilon^{n'}_{\rho\sigma} \langle 0| a^{\rho\sigma}(k') a^{\dagger\mu\nu}(k) |0\rangle \\
\nonumber
& = & \sum_{n,n'} \int\ud[3]{k} S^n_i S_f^{n'} \epsilon^n_{\mu\nu}  \epsilon^{n'}_{\rho\sigma} (\Itensorup{\mu}{\nu}{\rho}{\sigma}) \\
& = & 2 \sum_n \int\ud[3]{k} S_i^n S_f^n
\end{IEEEeqnarray}
using the normalization of the polarization tensors. This combines with the contribution (2) to give
\begin{equation}
-\sum_n \int\ud[3]{k} (S_f^n - S_i^n)^2.
\end{equation}

\item
The scalar-scalar-graviton vertex, with scalar momentum $p$ flowing in and $p'$ flowing out, is
\begin{equation}
-\frac{i}{2}\big[p_\mu p'_\nu + p_\nu p'_\mu - (1/2)\eta_{\mu\nu}(p\cdot p'-m^2)\big] \rightarrow -ip_\mu p_\nu \quad\text{as}\quad p-p'\rightarrow 0.
\end{equation}
The scalar propogator is $i/(p^2-m^2)$.  An external graviton line has an associated factor
\begin{equation}
\frac{\epsilon^l_{\mu\nu}}{\sqrt{(2\pi)^3 2k^0}}.
\end{equation}
For diagram (b), the additional scalar propagator has momentum $p_f-k$, so the diagrammatic factor is
\begin{multline}
\sum_n \int\ud[3]{k} S_f^n \frac{\epsilon^n_{\mu\nu}}{\sqrt{(2\pi)^3 2k^0}} (\Itensorup{\mu}{\nu}{\rho}{\sigma}) 
\left(\frac{i}{(p_f+k)^2-m^2}\right) \\ \times \left(\frac{-i}{2}\right) \big(p_f^\rho(p_f+k)^\sigma + p_f^\sigma(p_f+k)^\rho -\eta^{\rho\sigma}(p_f\cdot(p_f+k) - m^2)\big),
\end{multline}
or, in the limit $k\rightarrow 0$,
\begin{equation}
\sum_n \int\ud[3]{k} S_f^n \frac{\epsilon^n_{\mu\nu}}{\sqrt{(2\pi)^3 2k^0}} \left(\frac{1}{2p_f\cdot k}\right)\left(\frac{4p_f^\mu p_f^\nu}{2}\right)
= 2 \sum_n \int\ud[3]{k} S_f^n S_f^n.
\end{equation}
Similarly, in diagram (c) the additional propagator has momentum $p_i-k$, giving a factor
\begin{equation}
-2 \sum_n \int\ud[3]{k} S_f^n S_i^n.
\end{equation}
In diagram (d) the additional propagator has momentum $p_f-k$, giving
\begin{equation}
-2 \sum_n \int\ud[3]{k} S_i^n S_f^n,
\end{equation}
and in diagram (e) the additional progagator has momentum $p_i+k$, giving
\begin{equation}
2 \sum_n \int\ud[3]{k} S_i^n S_i^n.
\end{equation}
These factors from these four diagrams sum to
\begin{equation}
2\sum_n\int\ud[3]{k} (S^n_f - S^n_i)^2.
\end{equation}

\end{enumerate}

The total factor from real soft gravitons is then
\begin{IEEEeqnarray}{l}
\int\ud[3]{k}\sum_n (S_f^n - S_i^n)^2 \\ 
\nonumber
= \sum_n\int\ud[3]{k} \frac{1}{8k^0 (2\pi)^3} \left(\frac{p_f^\mu p_f^\nu \epsilon^n_{\mu\nu}}{p_f\cdot k} - \frac{p_i^\mu p_i^\nu \epsilon^n_{\mu\nu}}{p_i\cdot k}\right)^2 \\
\nonumber
= \int\!\frac{\mathrm{d}^3k}{k^0} \frac{1}{64\pi^3} \left(\frac{p_f^\mu p_f^\nu}{p_f\cdot k} - \frac{p_i^\mu p_i^\nu}{p_i\cdot k}\right)\left(\frac{p_f^\rho p_f^\sigma}{p_f\cdot k} - \frac{p_i^\rho p_i^\sigma}{p_i\cdot k}\right) \frac{1}{2}(\Itensordown{\mu}{\nu}{\rho}{\sigma}) \\
\nonumber
= \int\!\frac{\mathrm{d}^3k}{k^0} \frac{1}{64\pi^3} \left[\frac{p_f^4 - (1/2)p_f^4}{(p_f\cdot k)^2} + \frac{p_i^4 - (1/2)p_i^4}{(p_i\cdot k)^2} - 2\left(\frac{(p_f \cdot p_i)^2 - (1/2)(p_f)^2(p_i)^2}{(p_f\cdot k)(p_\cdot k)}\right)\right].
\end{IEEEeqnarray}
In the third line we have used the graviton polarization sum formula, which also contains terms proportional to the graviton momentum $k$.  But Weinberg shows in \cite{Weinberg}, using momentum conservation, that those terms do not contribute. This exactly cancels the contribution \eqref{virtualB} from virtual soft gravitons.

Before discussing the all-orders cancellation, we would like to comment on a rather unusual feature of the above discussion: the need to include the disconnected diagrams to exhibit the cancellation of the infrared divergences. In a related context, this was first emphasized in the classic paper of Lee and Nauenberg \cite{LN}, where it was pointed out that in general, in order to cancel infrared divergences the sum of transition probabilities over both initial and final state degenerate states must be carried out. It was noted there that including the initial state absorption contribution requires also the inclusion of disconnected diagrams. Additional confirmation that they play a necessary role was provided in \cite{ASZ}, where it was pointed out that the disconnected diagrams can be factored out, as is usually assumed, in transition probabilities, but the Feynman rules must be accordingly modified for the virtual diagrams and the real emissions. 

In the Faddeev-Kulish approach, the disconnected diagrams arise because the coherent states do not have a definite particle number. This is not special to the case of quantum gravity but was already used in Chung's analysis for QED \cite{Chung}.

	\subsection{Cancellation at all orders}
	\label{sec:allorders}

We now need to consider the all-orders infrared contributions to the matrix element.  The relevant factors are:
\begin{enumerate}
\item
As computed by Weinberg, soft gravitons in loops contribute a real exponential factor
\begin{equation}
\hspace{-.4in}\exp\left\{\frac{1}{2}\int\ud[4]{k}\left[\real B(k)\right]\right\} = \exp\left\{-\frac{1}{64\pi^3}\sum_{n,m}\int\!\frac{\mathrm{d}^3k}{k^0}\frac{\eta_n\eta_m[(p_n\cdot p_m)^2-(1/2)m_n^2 m_m^2]}{(p_n\cdot k)(p_m\cdot k)}\right\}.
\end{equation}

\item
The normalization of the initial and final states contributes
\begin{equation}
\exp\left\{-\sum_n\int\ud[3]{k}|S_i^n|^2\right\}\exp\left\{-\sum_n\int\ud[3]{k}|S_f^n|^2\right\}.
\end{equation}

\item
A diagram may contain some number $l$ of disconnected (noninteracting) graviton lines.  These contribute a factor
\begin{equation}
l!\left[2\sum_n\int\ud[3]{k}S_i^n S_f^n\right]^l,
\end{equation}
where $l!$ is the number of ways of pairing the $l$ incoming noninteracting gravitons with the $l$ outgoing noninteracting gravitons.

\item
Each initial state graviton connected to the body of the diagram contributes a factor
\begin{equation}
-2\sum_n\int\ud[3]{k} S_i^n(S^n+\xi(k)),
\end{equation}
where $\xi(k)$ is such that $S_i^n \xi(k)$ has an IR-convergent integral.  The initial state contains terms with an arbitrary number of gravitons, so there may be any number $m$ of attached incoming graviton lines.  They give an overall factor
\begin{equation}
\left[-2\sum_n\int\ud[3]{k} S_i^n(S^n+\xi(k))\right]^{m},
\end{equation}
which can be written, using the binomial expansion and interchanging the dummy variables $k$ at will between terms, as
\begin{equation}
\left(\prod_{r=1}^{m}(-2)\sum_{n_r}\int\ud[3]{k_r}S_i^{n_r}\right)\sum_{j=0}^{m}\frac{m!}{j!(m-j)!}\left[\prod_{i=1}^{j} S^{n_{i}}(k_{i})\right]\xi_{m-j}(k_{j+1}\dots k_m),
\end{equation}
where
\begin{equation}
\xi_{m-j}(k_{j+1}\dots k_m) = \prod_{t=j+1}^{m} \xi(k_{t})
\end{equation}
is IR finite. Similarly, $m'$ attached outgoing gravitons contribute a factor
\begin{equation}
\left(\prod_{r'=1}^{m'}(2)\sum_{n_{r'}}\int\ud[3]{k'_{r'}}S_f^{n_{r'}}\right)\sum_{j'=0}^{m'}\frac{m'!}{j'!(m'-j')!}\left[\prod_{i'=1}^{j'} S^{n_{i'}}(k'_{i'})\right]\xi'_{m'-j'}(k'_{j'+1}\dots k'_{m'}).
\end{equation}

\item
In summing over contributions from various numbers of initial and final state gravitons, some combinatoric factors occur:
\begin{enumerate}
\item Diagrams with $m$ interacting and $l$ noninteracting initial-state gravitons are produced by the $(m+l)$th term in the expansion of the exponential of creation operators.  So the contributions from these diagrams carry a factor $1/(m+l)!$.  Similarly, the expansion of the final state produces a factor $1/(m'+l)!$. 
\item The number of ways in which $m+l$ incoming gravitons can be divided into $m$ interacting and $l$ noninteracting is $(m+l)!/m!l!$.  Similarly, the number of ways in which $m'+l$ outgoing gravitons can be divided into $m'$ interacting and $l$ noninteracting is $(m'+l)!/m'!l!$.
\end{enumerate}
Altogether these combinatorics produce a factor
\begin{equation}
\frac{1}{m!l!}\frac{1}{m'!l!}.
\end{equation}
\end{enumerate}

Summing the external-graviton contributions (3) and (4), with the combinatoric factor (5), over all possible values of $m$,$m'$,$l$, gives
\begin{multline}
\sum_{l=0}^\infty\frac{1}{l!} \left[2\sum_n\int\ud[3]{k}S_i^n S_f^n\right]^l \\ \times
\sum_{m=0}^\infty \sum_{j=0}^{m}\frac{1}{j!(m-j)!} \left(\prod_{r=1}^{m}(-2)\sum_{n_r}\int\ud[3]{k_r}S_i^{n_r}\right)\left[\prod_{i=1}^{j} S^{n_{i}}(k_{i})\right]\xi_{m-j}(k_{j+1}\dots k_m) \\ \times
\sum_{m'=0}^\infty \sum_{j'=0}^{m'}\frac{1}{j'!(m'-j')!}  \left(\prod_{r'=1}^{m'}(2)\sum_{n_{r'}}\int\ud[3]{k'_{r'}}S_f^{n_{r'}}\right)\left[\prod_{i'=1}^{j'} S^{n_{i'}}(k'_{i'})\right]\xi'_{m'-j'}(k'_{j'+1}\dots k'_{m'}).
\end{multline}
The first factor exponentiates immediately.  In the second factor, we can exchange the order of summation using $j\leq m \Leftrightarrow m\geq j$ and then reindex $w=m-j$ to obtain
\begin{equation}
\sum_{j=0}^\infty \sum_{w=0}^{\infty}\frac{1}{j!w!} \left(\prod_{r=1}^{w+j}(-2)\sum_{n_r}\int\ud[3]{k_r}S_i^{n_r}\right)\left[\prod_{i=1}^{j} S^{n_{i}}(k_{i})\right]\xi_{w}(k_{j+1}\dots k_{w+j}).
\end{equation}
Within each term, the integrals factor into a part dependent only on the first $j$ momenta and one dependent only on the last $w$ momenta:
\begin{equation}
\sum_{j=0}^\infty \sum_{w=0}^{\infty}\frac{1}{j!w!} \Bigg(\prod_{r=1}^{j}(-2)\sum_{n_r}\int\ud[3]{k_r}S_i^{n_r} S^{n_{r}}(k_{r})\Bigg) \Bigg(\prod_{r=j+1}^{w+j}(-2)\sum_{n_r}\int\ud[3]{k_r}S_i^{n_r} \xi_{w}(k_{j+1}\dots k_{w+j})\Bigg).
\end{equation}
Another shift of index eliminates the apparent $j$-dependence of the last factor, so the sum over $j$ exponentiates:
\begin{equation}
\exp\left\{-2\sum_{n}\int\ud[3]{k}S_i^{n} S^{n}(k)\right\} \sum_{w=0}^{\infty}\frac{1}{w!} \left(\prod_{r=1}^{w}(-2)\sum_{n_r}\int\ud[3]{k_r}S_i^{n_r} \xi_{w}(k_{1}\dots k_{w})\right).
\end{equation}
Since $\xi_w$ is IR-finite, this isolates the IR divergence in the exponential.  We can do the same for the final-state factor, and then combining all the contributions (1)-(5) gives
\begin{multline}
\exp\left\{\frac{1}{2}\int\ud[4]{k}\left[\real B(k)\right]\right\}
\exp\left\{-\sum_n\int\ud[3]{k}|S_i^n|^2\right\}\exp\left\{-\sum_n\int\ud[3]{k}|S_f^n|^2\right\}
\\ \times
\exp\left\{2\sum_n\int\ud[3]{k}S_i^n S_f^n\right\}
\exp\left\{-2\sum_{n}\int\ud[3]{k}S_i^{n} S^{n}(k)\right\}
\exp\left\{2\sum_{n}\int\ud[3]{k}S_f^{n} S^{n}(k)\right\} M,
\end{multline}
where $M$ contains the IR-finite contributions (involving $\xi$) from initial- and final-state interacting gravitons.  Combining the exponents, we have
\begin{equation}
\exp\left\{\frac{1}{2}\int\ud[4]{k}\left[\real B(k)\right] +\sum_n\int\ud[3]{k}(S_f^n-S_i^n)^2 \right\} M.
\label{factor}
\end{equation}
The argument of the exponential is exactly the one-loop result, which was previously shown to vanish.
 
 \section{Infrared divergences in exclusive gravitational scattering amplitudes}
\label{sec:scattering}

In the previous section, we showed the cancellation of infrared divergences in a matrix element computed in the space of asymptotic states $\Hilbert_{as}$. It is worth emphasizing that within the Faddeev-Kulish framework, the soft and hard parts of an amplitude factor explicitly. This was seen directly for the example of potential scattering --- see eq. \eqref{factor}. We also showed in that case that the all-orders soft divergence can be obtained by exponentiating the one-loop divergence. This factorization is more straightforward in gravitational scattering than in gauge theories due to the absence of collinear divergences. 

In this section, using the Faddeev-Kulish approach, we will calculate the soft contributions to an exclusive gravitational scattering amplitude to all orders. Our approach is a generalization to gravitation of the calculation of the electron form factor using the QED asymptotic operator in \cite{Dahmen}. In order to remove any soft graviton radiation, we will need to impose constraints on the states. We treat the case of a four scalar scattering amplitude, though it will become clear that the same method applies for an arbitrary number of particles. In fact, our result can be expressed in terms of the expectation value of a product of gravitational Wilson line operators, in agreement with \cite{Schnitzer1, Schnitzer2} in which such a connection was conjectured.

Recall that the space of asymptotic states is given by 
\begin{equation}
\Hilbert_{as}=Z(t) \Hilbert_{Fock}=e^{i\Phi(t)}e^{R(t)}\Hilbert_{Fock}.
\end{equation}
Consider a four scalar exclusive scattering process, with all momenta taken as incoming. The soft part of this amplitude can be written as
\begin{align}
&\lim_{t\rightarrow\infty}\bra{\phi(p_{1})\phi(p_{2})\phi(p_{3})\phi(p_{4})}Z(t)\ket{0},
\end{align}
where $Z(t)$ is the asymptotic operator in the interaction representation, which is diagonal with respect to the momenta of the hard particles. In order to evaluate this matrix element, we project the operator $Z(t)$ into the four-scalar subspace:

\begin{equation}
\lim_{t\rightarrow\infty}\bra{\phi(p_{1})\phi(p_{2})\phi(p_{3})\phi(p_{4})}Z(t)\ket{0}
\nonumber = \lim_{t\rightarrow\infty}\bra{0}Z_{\phi}(t,p_{1},p_{2},p_{3},p_{4})\ket{0},
\end{equation}
where $Z_{\phi}$ is the asymptotic operator (in the interaction representation) with $V^{I}_{as}$ replaced by its projection $V_{\phi}$ onto the four-scalar subspace. In order to obtain an explicit form for $V_{\phi}$, it will be useful to express the asymptotic potential as in section \ref{sec:asymptotic}:
\begin{align}
V_{as}^{I}(t)=-\int \ud[3]{x}\ud[3]{p}\rho(p)\frac{p_{\mu}p_{\nu}}{2\omega}\delta^{3}(\vec{x}-t\vec{p}/\omega)h^{\mu\nu}(t,\vec{x}).
\end{align}
It is now easy to see that the projection of this operator onto the four scalar subspace has the form
\begin{IEEEeqnarray}{rCl}
V_{\phi}(t,p_{1},p_{2},p_{3},p_{4})&=&-\frac{p^{\mu}_{1}p^{\nu}_{1}h_{\mu\nu}(p_1 t / \omega_1)}{2\omega_{1}}-\frac{p^{\mu}_{2}p^{\nu}_{2}h_{\mu\nu}(p_2 t / \omega_2)}{2\omega_{2}}
\nonumber \\
&&-\frac{p^{\mu}_{3}p^{\nu}_{3}h_{\mu\nu}(p_3 t / \omega_3)}{2\omega_{3}}-\frac{p^{\mu}_{4}p^{\nu}_{4}h_{\mu\nu}(p_4 t / \omega_4)}{2\omega_{4}}
\nonumber \\
&\equiv&-G(\alpha)-G(\beta)-G(\eta)-G(\xi),
\end{IEEEeqnarray}
where $\alpha=p_1 t / \omega_1$, $\beta=p_2 t / \omega_2$, $\eta=p_3 t / \omega_3$, and $\xi=p_4 t / \omega_4$. Note that $Z_{\phi}$ will have two parts, one corresponding to the phase operator \eqref{Phi} and the other corresponding to the operator \eqref{R}. 

It is useful to define the commutator function
\begin{equation}
d(\alpha-\beta)\equiv [G(\alpha),G(\beta)]=i\left(\frac{2(p_{1}\cdot p_{2})^{2}-m^{4}}{4\omega_{1}\omega_{2}}\right)D(\alpha-\beta).
\end{equation}
Using this commutator function, we can write the projection of $i\Phi$ as
\begin{IEEEeqnarray}{Clrrrr}
& \IEEEeqnarraymulticol{5}{l}{
-\frac{1}{2}\int^{t}\ud{t_2}\int^{t_2}\ud{t_1} [V_{\phi}(t_{2},p_{1},p_{2},p_{3},p_{4}),V_{\phi}(t_{1},p_{1},p_{2},p_{3},p_{4})]}
\nonumber \\
= & -\frac{1}{2}\int^{t}\ud{t_{2}}\int^{t_{2}}\ud{t_{1}} & \big[\phantom{+}d(\alpha_{2}-\alpha_{1})&+d(\alpha_{2}-\beta_{1})&+d(\alpha_{2}-\eta_{1})&+d(\alpha_{2}-\xi_{1}) \phantom{\big]}
\nonumber \\
&& +\:d(\beta_{2}-\alpha_{1})&+d(\beta_{2}-\beta_{1})&+d(\beta_{2}-\eta_{1})&+d(\beta_{2}-\xi_{1}) \phantom{\big]}
\nonumber \\
&& +\:d(\eta_{2}-\alpha_{1})&+d(\eta_{2}-\beta_{1})&+d(\eta_{2}-\eta_{1})&+d(\eta_{2}-\xi_{1}) \phantom{\big]}
\nonumber \\
&& +\:d(\xi_{2}-\alpha_{1})&+d(\xi_{2}-\beta_{1})&+d(\xi_{2}-\eta_{1})&+d(\xi_{2}-\xi_{1})\big]
\nonumber \\
\equiv &
\IEEEeqnarraymulticol{5}{l}{
-\frac{1}{2}\int^{t}\ud{t_{2}}\int^{t_{2}}\ud{t_{1}}\sum_{A,B}d(A_{2}-B_{1}),}
\label{partone}
\end{IEEEeqnarray}
where the subscripts are such that $\beta_{1}=p_2 t_1 / \omega_2$ and the sum over $A, B$ is over all possible values in the set $\{\alpha,\beta,\eta,\xi \}$. Since the projection of the phase operator is just a c-number, we can pull it out of the expectation value. 

If we split up the function $G$ into its positive and negative frequency parts $G^{(+)}$ and $G^{(-)}$ and define another commutator function
\begin{equation}
d^{(+)}(\alpha-\beta)\equiv [G^{(-)}(\alpha),G^{(+)}(\beta)]=i\left(\frac{2(p_{1}\cdot p_{2})^{2}-m^{4}}{4\omega_{1}\omega_{2}}\right)D^{(+)}(\alpha-\beta),
\end{equation}
then we can write the expectation value of the projection of $e^{R}$ as
\begin{IEEEeqnarray}{Clr}
& \IEEEeqnarraymulticol{2}{l}{
\bra{0}\exp\left\{-i\int^{t}\ud{t_{1}}V_{\phi}(t_{1},p_{1},p_{2},p_{3},p_{4})\right\}\ket{0}}
\nonumber \\
= &\exp\Bigg\{\frac{1}{2}\int^{t}\ud{t_{1}}\int^{t}\ud{t_{2}}&\Big[G^{(-)}(\alpha_{2})+G^{(-)}(\beta_{2})+G^{(-)}(\eta_{2})+G^{(-)}(\xi_{2})\phantom{\Big]\Bigg\}}
\nonumber \\
&&G^{(+)}(\alpha_{1})+G^{(+)}(\beta_{1})+G^{(+)}(\eta_{1})+G^{(+)}(\xi_{1})\Big]\Bigg\}
\nonumber \\
=& \IEEEeqnarraymulticol{2}{l}{
\exp\left\{\frac{1}{2}\int^{t}\ud{t_{1}}\int^{t}dt_{2}\sum_{A,B}d^{(+)}(A_{2}-B_{1})\right\}},
\label{parttwo}
\end{IEEEeqnarray}
where the Baker-Campbell-Hausdorff formula and the constraint $G^{(+)}\ket{0}=\bra{0}G^{(-)}=0$ have been used in the second line. The constraint serves to restrict us to the exclusive amplitude, i.e. the case of no real external soft gravitons. 

Combining \eqref{partone} and \eqref{parttwo}, we arrive at
\begin{align}
\lim_{t\rightarrow\infty}\bra{\phi(p_{1})\phi(p_{2})\phi(p_{3})\phi(p_{4})}Z(t)\ket{0}=e^{-I_{1}-I_{2}},
\end{align}
where
\begin{IEEEeqnarray}{rClr}
I_{1} &=& \frac{1}{2}\int^{\infty}\ud{t_{1}}\int^{\infty}\ud{t_{2}}\sum_{A\ne B}&\left(\theta(t_{2}-t_{1})d(A_{2}-B_{1})-d^{(+)}(A_{2}-B_{1})\right),
\\
I_{2}&=&\frac{1}{2}\int^{\infty}\ud{t_{1}}\int^{\infty}\ud{t_{2}}\sum_{A}&\left(\theta(t_{2}-t_{1})d(A_{2}-A_{1})-d^{(+)}(A_{2}-A_{1})\right).
\end{IEEEeqnarray}
Noting that
\begin{align}
\theta(t_{2}-t_{1})d(\alpha_{2}-\beta_{1})-d^{(+)}(\alpha_{2}-\beta_{1})=\frac{i}{(2\pi)^{4}}\frac{2(p_{1}\cdot p_{2})^{2}-m^{4}}{4\omega_{1}\omega_{2}}\int \ud[4]{k}\frac{e^{-ik(\alpha_{2}-\beta_{1})}}{k^{2}-i\epsilon},
\end{align}
we can write:
\begin{IEEEeqnarray}{rCl}
I_{1}&=&\frac{i}{4(2\pi)^{4}}\sum_{i<j}\int \ud[4]{k} \frac{2(p_{i}\cdot p_{j})^{2}-m^{4}}{(k^{2}-i\epsilon)(k\cdot p_{i})(k \cdot p_{j})}, \\
I_{2}&=&\frac{i}{8(2\pi)^{4}}\sum_{i}\int \ud[4]{k} \frac{m^{4}}{k^{2}}\frac{1}{(k\cdot p_{i})^{2}}.
\end{IEEEeqnarray}
This is our final result for the soft part of the four-particle scattering amplitude. To compare with the results of \cite{Schnitzer1}, we now take the massless limit for the external scalars. In this limit $I_{2}$ vanishes and we are left with
\begin{align}
I_{1}\Big|_{m=0}=\frac{i}{4}(\kappa\mu^{\epsilon})^{2}\sum_{i<j}\int \udpi[d]{k} \frac{2(p_{i}\cdot p_{j})^{2}}{(k^{2}-i\epsilon)(k\cdot p_{i})(k \cdot p_{j})},
\label{SIntegral}
\end{align}
where we have written the integral in $d=4-2\epsilon$ dimensions and restored dependence on the gravitational constant $\kappa=\sqrt{16\pi G}$, and where $\mu$ is an arbitrary mass scale introduced in order to keep $I_{1}|_{m=0}$ dimensionless. This is exactly the one-loop soft function obtained in \cite{Schnitzer1}. There, the soft function for n graviton scattering is hypothesized to be
\begin{align}
S_n = \bra{0}\prod_{i=1}^n\Phi_{p_i}(0, \infty)\ket{0},
\end{align}
where the gravitational Wilson line operator is
\begin{align}
\Phi_p(a, b) = \mathrm{P} \exp\left\{i\frac{\kappa}{2}\int_a^b \ud{s} p^{\mu}p^{\nu} h_{\mu\nu}(sp)\right\}.
\end{align}

Our approach not only picks out the gravitational Wilson line operator of \cite{Schnitzer1} as the correct representation of the soft function but also demonstrates straightforwardly the exponentiation of the one loop result, eliminating the need to perform an order by order investigation in perturbation theory. 

The integral \eqref{SIntegral} actually vanishes due to a cancellation between the UV and IR divergences. However, the UV divergence is artificial and should be compensated by a counterterm. It arises because we took the graviton momentum $k$ to be small when we constructed the asymptotic operator; the large-$k$ regime here is not meaningful. Using dimensional regularization, it was shown in \cite{Schnitzer1} that \eqref{SIntegral} takes the form
\begin{align}
I_{1}\Big|_{m=0}=-\frac{1}{2}(\kappa\mu^{\epsilon})^{2}\frac{\Gamma(1+\epsilon)}{(4\pi)^{d/2}}\sum_{i<j}\frac{(2p_{i}\cdot p_{j})^{1-\epsilon}}{\epsilon^{2}}.
\end{align}
If we define $\lambda=\frac{1}{2}(\kappa)^{2}(4\pi e^{-\gamma})^{\epsilon}$ and use momentum conservation $\sum_{i<j}p_{i}\cdot p_{j}=0$, we arrive at the result
\begin{align}
-I_{1}\Big|_{m=0}=\frac{\lambda}{(4\pi)^{2}\epsilon}\sum_{i<j}(-2p_{i}\cdot p_{j})\log\left(\frac{2p_{i}\cdot p_{j}}{\mu^{2}}\right).
\label{n=8}
\end{align}

The absence of ${1 / \epsilon^2}$ divergences in the massless limit signals the cancellation of the collinear singularities. While we have treated the four-scalar case explicitly, it is clear that this result applies for an arbitrary number of external particles, since momentum conservation guarantees $\sum_{i<j}p_{i}\cdot p_{j}=0$ for any number of external momenta.

\section{Conclusions}

We have constructed a gravitational S matrix $S_A$ and a space of asymptotic states analogous to those defined by Faddeev and Kulish in QED. We find that infrared divergences vanish in matrix elements of $S_A$ without the need to sum over cross sections. As an application of the asymptotic state formalism we have also calculated the virtual soft corrections to a $2 \rightarrow 2$ hard scattering process and related these to a gravitational Wilson line contribution.
While we consider the simple case of scalar matter content, our treatment is in fact general, since soft gravitons are insensitive to spin. 

The construction of coherent asymptotic states was made possible by the surprisingly simple IR structure of perturbative gravity: divergences only occur in subdiagrams in which no vertex connects more than one soft graviton line. As a result, we were able to restrict our consideration to linearized interactions, and the coherent states of quantum electrodynamics and perturbative quantum gravity turned out to to be quite similar. The key difference between the two theories is that the spin-2 graviton has derivative couplings, which, through momentum conservation, prevent the appearance of collinear divergences.

By providing a well-defined S matrix, this work establishes a firm foundation for perturbative studies of quantum gravity. Since no infrared divergences are associated with massless fermions including gravitini, or with massless scalars, the long distance  behavior of supergravity is the same as that of gravity and massless vectors. For the case of high energy small angle scattering of gravitons this was observed in \cite{Giddings2}. Thus, our results are also applicable to ($\mathcal{N}=8)$ supergravity, in which IR divergences have until now been handled only at the level of inclusive cross sections \cite{Zhiboedov}. In fact, eq. \eqref{n=8} is the infrared divergent part of the four graviton scattering amplitude in $\mathcal{N}=8$ supergravity as calculated in \cite{Bern}. Formulations of the Fadeev-Kulish procedure in terms of surface terms at infinity \cite{Zwanziger} also suggest a possible application to holographic renormalization \cite{holography, holography2}.

Our discussion of the asymptotic Hamiltonian and the state space naturally ties in to the question of asymptotic symmetries at future and past null infinities. In \cite{BMS, Sachs} aspects of this asymptotic symmetry group, called the BMS group, were explored. Clearly, the infrared sectors of the massless fields must be connected to the representations of the BMS group. In a fascinating recent paper \cite{Strominger}, an extension of the BMS algebra was discussed which allows for analytic singularities and a connection with the Weinberg soft graviton theorems \cite{Weinberg} was established. These developments are currently under investigation in the context of the Faddeev-Kulish \cite{FK} approach discussed here.

\begin{appendices}
\section{Gravitational `Coulomb' phase in dimensional regularization} \label{Appendix A}
In this appendix we provide a proof of eq. \eqref{gravphi}, which expresses the divergent gravitational phase factor in dimensional regularization. Weinberg \cite{Weinberg} shows (his eq. (2.19)) that the contribution of internal soft graviton lines to an S-matrix element is a multiplicative factor
\begin{equation}
\frac{S_{\beta\alpha}}{S_{\beta\alpha}^0}=\exp\left\{\frac{1}{2}\int^\Lambda \ud[4]{q} B(q) \right\},
\end{equation}
where
\begin{equation}
B(q) = \frac{8\pi G i}{(2\pi)^4} \sum_{nm} \frac{[(p_n\cdot p_m)^2 - \frac{1}{2}m_n^2 m_m^2] \eta_n\eta_m}{(q^2-i\epsilon)(p_n\cdot q-i\eta_n\epsilon)(p_m\cdot q + i\eta_m\epsilon)}.
\end{equation}
The sum runs over pairs $n,m$ of initial and/or final state particles, with $\eta=+1$ for outgoing particles and $\eta=-1$ for incoming particles.  We've kept Weinberg's UV cutoff $\Lambda$, which separates `soft' from `hard' virtual gravitons, but removed the IR cutoff or photon mass $\lambda$; we use dimensional regularization for IR divergences instead.  We can write
\begin{equation}
\frac{S_{\beta\alpha}}{S_{\beta\alpha}^0}=\exp\left\{\frac{4\pi G}{(2\pi)^4}\sum_{nm} [(p_n\cdot p_m)^2 - \frac{1}{2}m_n^2 m_m^2] \eta_n\eta_m J_{nm} \right\},
\end{equation}
\begin{equation}
J_{nm} = i\int^\Lambda\ud[4]{q} \frac{1}{(q^2-i\epsilon)(p_n\cdot q-i\eta_n\epsilon)(p_m\cdot q + i\eta_m\epsilon)}.
\end{equation}

This is now essentially the same integral as Weinberg eq. (5.5).  The $q^0$ contour integration is the same: if $\eta_n, \eta_m$ are opposite, then the poles $q^0=\vec{v_n}\cdot\vec{q}-i\eta_n\epsilon$, $q^0 = \vec{v_m}\cdot\vec{q}+i\eta_m\epsilon$ lie on the same side of the real axis and may both be avoided.  Then $J_{nm}$ is real.  But if particles $n$ and $m$ are either both incoming or both outgoing, then the poles lie on opposite sides of the real axis and cannot both be avoided.  Take the case $\eta_n=\eta_m=+1$; the other is the same.  Closing the contour in the upper half-plane, we pick up contributions from the poles at $-|\vec{q}|+i\epsilon$ and $\vec{v_m}\cdot q +i\epsilon$, where as in Weinberg $\vec{v}=\vec{p}/E$.  The residue theorem gives
\begin{multline}
J_{nm} = i\int^\Lambda\ud[3]{q} (2\pi i) \Bigg[\frac{1}{2|\vec{q}|(\vec{q}\cdot\vec{p_n}-|\vec{q}|E_n)(\vec{q}\cdot\vec{p_m}-|\vec{q}|E_m)} \\+ \frac{1}{(|\vec{q}|^2-(\vec{v_m}\cdot\vec{q})^2)(\vec{q}\cdot\vec{p_n}-E_n\vec{v_m}\cdot\vec{q}-i\epsilon)(E_m)} \Bigg].
\end{multline}
The $i\epsilon$-prescription has been retained where it will become important later.

The first term is again the real part; the second will give the divergent phase.  It can be written
\begin{equation}
-\frac{2\pi}{E_n E_m}\int^\Lambda\ud[3]{q} \frac{1}{|\vec{q}|^2(1-v_m^2\cos^2\theta)[|\vec{q}||\vec{v_n}-\vec{v_m}|(\cos\theta \cos\beta - \sin\theta\cos\phi\sin\beta)-i\epsilon]},
\end{equation}
where $\beta$ is the angle between $\vec{v_m}$ and $\vec{v_m}$.  Since the result must be Lorentz-invariant, we can evaluate it in the center-of-momentum frame of particles $n$ and $m$.  Then the velocities are back-to-back, so $\beta=\pi$.  That leaves
\begin{equation}
\frac{2\pi}{E_n E_m |\vec{v_n}-\vec{v_m}|}\int^\Lambda\ud[3]{q} \frac{1}{|\vec{q}|^3(1-v_m^2\cos^2\theta)(\cos\theta-i\epsilon)}.
\end{equation}
Continue the integral to $n-1$ spatial dimensions, using the prescription for spherical coordinates given in \cite{Marciano-Sirlin}:
\begin{equation}
\frac{(2\pi)^4}{E_n E_m |\vec{v_n}-\vec{v_m}|}\int^\Lambda\udpi[n-1]{q} \frac{1}{|\vec{q}|^3(1-v_m^2\cos^2\theta)(\cos\theta-i\epsilon)}.
\end{equation}

Choosing $\vec{v_m}$ to be along the first axis, the integrand is independent of the angles $\theta_2\dots\theta_{n-2}$.  Carrying out those integrals yields an overall factor:
\begin{equation}
\frac{(2\pi)^{5-n}}{E_n E_m |\vec{v_n}-\vec{v_m}|} \frac{2 \pi^{(n/2-1)}}{\Gamma(n/2-1)} \int_0^\Lambda\ud{q} q^{n-2} \frac{1}{q^3} \int_0^\pi \ud{\theta} \sin^{n-3}\theta \frac{1}{(1-v_m^2\cos^2\theta)(\cos\theta-i\epsilon)}.
\end{equation}
Substituting $y=\cos\theta$, we have
\begin{equation}
\int_{-1}^1\ud{y} \frac{(1-y^2)^{(n/2-1)}}{(1-v_m^2 y^2)} \frac{1}{(y-i\epsilon)}.
\end{equation}
The real part is odd and integrates to zero, but the imaginary part $i\pi\delta(y)$ of the pole integrates to $i\pi$.

Now do the integral over $q$.  To regulate the IR divergence, we should take $n>4$, so
\begin{equation}
\int_0^\Lambda\ud{q} q^{n-5} = \frac{\Lambda^{n-4}}{n-4}.
\end{equation}
Altogether we have
\begin{equation}
\frac{2^{6-n}\pi^{(5-n/2)}i}{E_n E_m |\vec{v_n}-\vec{v_m}|\Gamma(n/2-1)} \frac{\Lambda^{n-4}}{n-4} = \frac{2\pi^3 i}{E_n E_m |\vec{v_n}-\vec{v_m}|}\frac{2}{\epsilon} + \text{IR finite},
\end{equation}
where now $\epsilon=n-4$.  Note that terms proportional to $\ln(\Lambda)$ are finite because $\Lambda$ serves here not as a regulator of UV divergences (which should be handled by renormalization) but merely as a finite separator between soft and hard graviton momenta.  Restoring the kinematic factor to covariant form gives
\begin{equation}
\imaginary J_{nm} = \frac{2\pi^3}{((p_n\cdot p_m)^2-m_n^2 m_m^2)^{1/2}}\frac{2}{\epsilon} + \text{IR finite}.
\end{equation}
So the divergent phase associated with virtual soft gravitons is
\begin{equation}
\exp\left\{\frac{iG}{2}\sum_{nm}\frac{[2(p_n\cdot p_m)^2 - m_n^2 m_m^2]}{[(p_n\cdot p_m)^2 - m_n^2 m_m^2]^{1/2}}\frac{1}{\epsilon} \right\}.
\end{equation}
The sum here counts the pairs $(n,m)$ and $(m,n)$ separately, so each distinct pair of initial or final state particles contributes
\begin{equation}
\exp\left\{iG\frac{[2(p_n\cdot p_m)^2 - m_n^2 m_m^2]}{[(p_n\cdot p_m)^2 - m_n^2 m_m^2]^{1/2}}\frac{1}{\epsilon} \right\},
\end{equation}
or, putting the kinematics in terms of the relative velocity $\beta = \sqrt{1-\frac{m_n^2 m_m^2}{(p_n\cdot p_m)^2}}$ and taking $16\pi G = 1$ and, the divergent phase is
\begin{equation}
\phi_{\text{grav}} = \frac{i}{16\pi}\frac{m_n m_m(1+\beta^2)}{\beta\sqrt{1-\beta^2}}\frac{1}{\epsilon}.
\end{equation}

\end{appendices}

\bibliographystyle{utphys}

\begin{thebibliography}{99}

\bibitem{Weinberg}
S.~Weinberg, \emph{Infrared Photons and Gravitons}, \href{http://prola.aps.org/abstract/PR/v140/i2B/pB516_1}{\emph{Phys. Rev.} {\bf 140} (1965), B516.}
	
\bibitem{Donoghue}
J.~F.~Donoghue and T.~Torma, \emph{Infrared behavior of graviton-graviton scattering}, \href{http://prd.aps.org/abstract/PRD/v60/i2/e024003}{\emph{Phys. Rev. D} {\bf 60} (1999), 024003} \href{http://arxiv.org/abs/hep-th/9901156v1}{[arXiv:hep-th/9901156]}.

\bibitem{BN}
F.~Bloch and A.~Nordsieck, \emph{Note on the radiation field of the electron}, \href{http://prola.aps.org/abstract/PR/v52/i2/p54_1}{{\emph{Phys. Rev.}{\bf 52} (1937), 54}}.

\bibitem{'tHooft}
G.~'t Hooft, \emph{The scattering matrix approach for the quantum black hole, an overview} \href{http://www.worldscientific.com/doi/abs/10.1142/S0217751X96002145}{\emph{Int. J. Mod. Phys. A}, {\bf 11} (1996), 4623} \href{http://arxiv.org/abs/gr-qc/9607022}{[arXiv:gr-qc/9607022]}.
arXiv:
\bibitem{Giddings1}
S.~B.~Giddings, \emph{The gravitational S matrix: Erice lectures}, \href{http://arxiv.org/abs/1105.2036}{[arXiv:1105.2036 [hep-th]]}.

\bibitem{amplitudes}
L.~J.~Dixon, \emph{Calculating scattering amplitudes efficiently}, \href{http://arxiv.org/abs/hep-ph/9601359}{[arXiv:hep-ph/9601359]}.

\bibitem{amplitudes2}
L.~J.~Dixon, \emph{Scattering amplitudes: the most perfect microscopic structures in the universe}, \href{http://iopscience.iop.org/1751-8121/44/45/454001}{{\emph{J. Phys. A} {\bf 44}, (2011), 454001}} \href{http://arxiv.org/abs/1105.0771}{[arXiv:1105.0771 [hep-th]]}.

\bibitem{amplitudes3}
H.~Elvang and Y.~-t.~Huang, \emph{Scattering Amplitudes},
  \href{http://arxiv.org/abs/1308.1697}{[arXiv:1308.1697 [hep-th]]}.

\bibitem{FK}
P.~P.~Kulish and L.~D.~Faddeev, \emph{Asymptotic conditions and infrared divergences in quantum electrodynamics}, \href{http://www.springerlink.com/content/l320g77qr47264gx/}{\emph{
  Theor. Math. Phys.} {\bf 4} (1965), 745}.

\bibitem{Contopanagos}
H.~F.~Contopanagos and M.~B.~Einhorn, \emph{Interpretation of the asymptotic S matrix for massless particles}, \href{http://prd.aps.org/abstract/PRD/v45/i4/p1291_1}{\emph{Phys. Rev. D} {\bf 45} (1992), 1291}.
	
\bibitem{Coll}
R.~Akhoury, R.~Saotome, and G.~Sterman, \emph{Collinear and soft divergences in perturbative quantum gravity}, \href{http://prd.aps.org/abstract/PRD/v84/i10/e104040}{\emph{Phys. Rev. B} {\bf 84} (2011), 104040} \href{http://arxiv.org/abs/1109.0270}{[arXiv:1109.0270 [hep-th]]}.

\bibitem{Chung}
V.~Chung, \emph{Infrared Divergence in Quantum Electrodynamics}, \href{http://prola.aps.org/abstract/PR/v140/i4B/pB1110_1}{\emph{
  Phys. Rev.} {\bf 140} (1965), B1110}.

\bibitem{Schnitzer1}
S.~G.~Naculich and H.~J.~Schnitzer, \emph{Eikonal methods applied to gravitational scattering amplitudes}, \href{http://www.springerlink.com/content/8v72wn7761618438/}{\emph{
  JHEP} {\bf 05} (2011), 87} \href{http://arxiv.org/abs/1101.1524}{[arXiv:1101.1524 [hep-th]]}.
  
\bibitem{Schnitzer2}
S.~Melville, S.~G.~Naculich, H.~J.~Schnitzer and C.~D.~White, 
 \emph{Wilson line approach to gravity in the high energy limit}
  \href{http://arxiv.org/abs/1306.6019}{[arXiv:1306.6019 [hep-th]]}.
  
  \bibitem{Beneke}
  M.~Beneke and G.~Kirilin, \emph{Soft-collinear gravity}, 
 \href{http://link.springer.com/article/10.1007/JHEP09\string%282012\string%29066}{{\emph{JHEP} {\bf 1209}, 066 (2012)}}
  \href{http://arxiv.org/abs/1207.4926}{[arXiv:1207.4926 [hep-ph]]}.
	
\bibitem{Weinberg2}
S.~Weinberg, \emph{Photons and Gravitons in S-Matrix Theory: Derivation of Charge Conservation and Equality of Gravitational and Inertial Mass}, \href{http://prola.aps.org/abstract/PR/v135/i4B/pB1049_1}{{\emph{Phys. Rev.} {\bf 135} (1964), B1049}}.
	
\bibitem{deWitt}
B.~S.~deWitt, \emph{Quantum Theory of Gravity. III. Applications of the Covariant Theory,} \href{http://prola.aps.org/abstract/PR/v162/i5/p1239_1}{\emph{Phys. Rev.} {\bf 162} (1967), 1239.}
	
\bibitem{Aichelburg-Sexl}
P.~C.~Aichelburg and R.~U.~Sexl, \emph{On the gravitational field of a massless particle}, \href{http://link.springer.com/article/10.1007/BF00758149\string#}{\emph{Gen. Rel. Grav.} {\bf 2} (1971), 303}.

\bibitem{WeinbergBook}
S.~Weinberg, \emph{The Quantum Theory of Fields, Vol. 1: Foundations}, Cambridge University Press, Cambridge U.K. (1995), pg. 67.

\bibitem{LN}T.~D.~Lee and M.~Nauenberg, 
 \emph{Degenerate Systems and Mass Singularities}, 
\href{http://prola.aps.org/abstract/PR/v133/i6B/pB1549_1}{{\emph{Phys. Rev.} {\bf 133} (1964), B1549}}.

\bibitem{ASZ}R.~Akhoury, M.~G.~Sotiropoulos and V.~I.~Zakharov, 
 \emph{The KLN theorem and soft radiation in gauge theories: Abelian case}, 
\href{http://prd.aps.org/abstract/PRD/v56/i1/p377_1}{{\emph{Phys. Rev. D} {\bf 56} (1997), 377}} \href{http://arxiv.org/abs/hep-ph/9702270}{[arXiv:hep-ph/9702270]}.

\bibitem{Gupta}
S.~Gupta, \emph{Quantization of Einstein's Gravitational Field: Linear Approximation}, \href{http://iopscience.iop.org/0370-1298/65/3/301}{\emph{Proc. Phys. Soc.} {\bf A 65} (1952), 161.}
 
\bibitem{Dahmen}
H.~D.~Dahmen, B.~Scholtz, and F.~Steiner, \emph{Infrared dynamics of quantum electrodynamics and the asymptotic behavior of the electron form factor}, \href{http://www.sciencedirect.com/science/article/pii/0550321382904047}{{\emph{Nucl. Phys.} {\bf B 202} (1982), 365.}}

\bibitem{Giddings2}
S.~ B.~ Giddings, M.~Schmidt-Sommerfeld, J.~R.~Andersen, \emph{High energy scattering in gravity and supergravity}, \href{http://prd.aps.org/abstract/PRD/v82/i10/e104022}{{\emph{Phys. Rev. D} {\bf 82} (2010) 104022}} \href{http://arxiv.org/abs/1005.5408}{[arXiv:1005.5408 [hep-th]]}.
	
\bibitem{Zhiboedov}
L.~V.~Bork, D.~I.~Kazakov, G.~S.~Vartanov, and A.~V.~Zhiboedov, \emph{Infrared finite observables in $N = 8$ supergravity}, \href{http://link.springer.com/article/10.1134\string%2FS0081543811010056}{\emph{Proc. Steklov Inst. Math.} {\bf 272} (2011), 39.}

\bibitem{Bern}
Z.~Bern, L.~J.~Dixon, D.~C.~Dunbar, M.~Perelstein and J.~S.~Rozowsky,
\emph{On the relationship between Yang-Mills theory and gravity and its implication for ultraviolet divergences}, \href{http://www.sciencedirect.com/science/article/pii/S0550321398004209}{{\emph{Nucl. Phys.} {\bf B 530} (1998), 401}} \href{http://arxiv.org/abs/hep-th/9802162}{[arXiv:hep-th/9802162]}.

\bibitem{Zwanziger}
J.~L.~Gervais and D.~Zwanziger, \emph{Derivation from first principles of the infrared structure of quantum electrodynamics}, \href{http://www.sciencedirect.com/science/article/pii/037026938090903X}{\emph{Phys. Lett.} {\bf B 94} (1980), 389.}

\bibitem{holography}
M.~Fukuma, S.~Matsuura, and T.~Sakai, \emph{Holographic Renormalization Group}, \href{http://ptp.ipap.jp/link?PTP/109/489}{\emph{Prog. Theor. Phys.} {\bf 109} (2003), 489.}

\bibitem{holography2}
I.~Papadimitriou, \emph{Holographic renormalization as a canonical transformation}, \href{http://link.springer.com/article/10.1007\string%2FJHEP11\string%282010\string%29014}{\emph{JHEP} {\bf 11} (2010), 014} \href{http://arxiv.org/abs/1007.4592}{[arXiv:1007.4592 [hep-th]]}.

\bibitem{BMS}
H.~Bondi, M.~G.~J.~van der Burg, A.~W.~K.~Metzner, \emph{Gravitational waves in general relativity VII. Waves from isolated axisymmetric systems}, \href{http://rspa.royalsocietypublishing.org/content/269/1336/21.abstract}{{\emph{Proc. Roy. Soc. Lond.} A {\bf 269} (1962), 21.}}

\bibitem{Sachs}
R.~K.~Sachs, \emph{Gravitational waves in general relativity VIII. Waves in asymptotically flat space-time}, \href{http://rspa.royalsocietypublishing.org/content/270/1340/103.abstract}{{\emph{Proc. Roy. Soc. Lond.} A {\bf 270} (1962), 103}.}

\bibitem{Strominger}
A.~Strominger, \emph{Asymtotic symmetries of Yang-Mills theory}, \href{http://arxiv.org/abs/1308.0589}{[arXiv: 1308.0589[hep-th]].}

\bibitem{Marciano-Sirlin}
W.~J.~Marciano and A.~Sirlin, \emph{Dimensional regularization of infrared divergences}, \href{http://www.sciencedirect.com/science/article/pii/0550321375905271}{\emph{Nucl. Phys.} {\bf B 88} (1975), 86.} 

\end{thebibliography}

\end{document}